\documentclass[preprintnumbers, floatfix,letterpaper,aps,prd,epsfig,nofootinbib,
twocolumn
]{revtex4-1}
\usepackage{bm,graphicx,dcolumn,epstopdf,epsf, latexsym,mathbbol, amssymb,amsmath,color,slashed, mathrsfs,mathcomp,simplewick}
\usepackage[center]{subfigure}
\usepackage{multirow}
\usepackage{makecell}
\usepackage[colorlinks,linkcolor=blue,citecolor=blue,urlcolor=blue]{hyperref}

\begin{document}
\allowdisplaybreaks
 \newcommand{\bq}{\begin{equation}}
 \newcommand{\eq}{\end{equation}}
 \newcommand{\bqn}{\begin{eqnarray}}
 \newcommand{\eqn}{\end{eqnarray}}
 \newcommand{\nb}{\nonumber}
 \newcommand{\lb}{\label}
 \newcommand{\f}{\frac}
 \newcommand{\p}{\partial}
\newcommand{\PRL}{Phys. Rev. Lett.}
\newcommand{\PLB}{Phys. Lett. B}
\newcommand{\PRD}{Phys. Rev. D}
\newcommand{\CQG}{Class. Quantum Grav.}
\newcommand{\JCAP}{J. Cosmol. Astropart. Phys.}
\newcommand{\JHEP}{J. High. Energy. Phys.}

\title{Shadows and deflection angle of charged and slowly rotating black holes in Einstein-\AE ther theory}

\author{Tao Zhu${}^{a, b}$}
\email{zhut05@zjut.edu.cn}

\author{Qiang Wu${}^{a}$}
\email{wuq@zjut.edu.cn; Corresponding author}

%

\author{Mubasher Jamil${}^{a,c}$ }
\email{mjamil@zjut.edu.cn}

\author{Kimet Jusufi${}^{d,e}$}
\email{kimet.jusufi@unite.edu.mk}

\affiliation{${}^{a}$ Institute for Theoretical Physics and Cosmology, Zhejiang University of Technology, Hangzhou, 310023, China
\\${}^{b}$ GCAP-CASPER, Physics Department, Baylor University, Waco, TX 76798-7316, USA}
\affiliation{${}^{c}$ School of Natural Sciences, National University of Sciences and Technology, Islamabad, 44000, Pakistan}

\affiliation{${}^{d}$Physics Department, State University of Tetovo, Ilinden Street nn, 1200,
Tetovo, North Macedonia}
\affiliation{${}^{e}$Institute of Physics, Faculty of Natural Sciences and Mathematics, Ss. Cyril
and Methodius University, Arhimedova 3, 1000 Skopje, North Macedonia}
\date{\today}

%
%

\begin{abstract}
In this paper, we study the shadow cast by two types of charged and slowly rotating black holes in Eisntein-\AE{}ther theory of gravity. This two types of black holes are corresponding to two specific combinations of the coupling constants of the \ae{}ther field, i.e., $c_{14}=0$ but $c_{123}\neq 0$ for the first type and $c_{123}=0$ for the second type, respectively. For both types of black holes, in addition to the mass and charge of the black holes, we show that the presence of the \ae{}ther field can also affect the size of the shadow. For the first type black hole, it is shown that the shadow size increases with the parameter $c_{13}$, while for the second type black hole, the shadow size still increases with $c_{13}$ but decreases with the parameter $c_{14}$. With these properties of these \ae{}ther parameters, we also discuss the observational constraints on these parameters by using the data of the first black hole image by the Event Horizon Telescope. In addition, we also explore the effect of the \ae{}ther field on the the deflection angle of light and the time delay by using the Gauss-Bonnet theorem. It is shown that, for a specific combination $c_{123} =0$, the deflection angle/time delay is slightly affected  by the \ae{}ther parameter $c_{13}$ at the leading order.
\end{abstract}


\maketitle

\section{Introduction}
\renewcommand{\theequation}{1.\arabic{equation}} \setcounter{equation}{0}

Very recently, the Event Horizon Telescope (EHT) Collaboration announced their first image concerning the detection of an event horizon of a supermassive black hole at the center of a neighboring elliptical M87 galaxy \cite{m87, Akiyama:2019eap}. The bright accretion disk surrounding the black hole appears distorted due to the phenomenon of gravitational lensing. The region of accretion disk behind the black hole also gets visible due to bending of light by black hole. This shadow image provides us a great opportunity to test the predictions of general relativity in the regime of strong gravity and improve our understanding about the geometrical structure of the event horizon of the black hole.  With this image, it is observed that the diameter of the center black hole shadow is $(42\pm 3) \; {\rm \mu as}$, which leads to a measurement of the center mass of $M=(6.5\pm 0.7)\times 10^9 M_{\odot}$ \cite{m87}. Future improvement of the observations, such as those together with the the Next Generation Very Large Array \cite{ngVLA}, the Thirty Meter Telescope \cite{TMT}, and BlackHoleCam \cite{BlackHoleCam}, can provide more precise and details information about the nature of the black hole spacetime in the regime of strong gravity. Importantly, these precise observations can also provide a significant way to distinguish or constrain black holes in different gravity theories.

Theoretically, the shadow cast by the black hole horizon is studied as null geodesics and the existence of a photon sphere or photon region (see \cite{Claudel:2000yi} and references therein for discussions about the geometric properties of photon sphere.). For a The incoming photons are trapped in an unstable circular orbit ($r=3M$ for a Schwarzschild black hole). Occasionally the photons are perturbed and diverted towards the observer. The shadow of a black hole is characterized by the celestial coordinates $\alpha$ and $\beta$ which are plotted for different values of the black hole parameters such as mass and spin. In literature, the theory of black hole shadows is well-developed and is under investigation for decades. Some notable results of black hole shadows are already discussed in the literature, to list a few: Schwarzschild black hole surrounded by a Bach-Weyl ring \cite{wang_shadow_2019}; Einstein-dilaton-Gauss-Bonnet black hole \cite{gb}; Konoplya-Zhidenko or more general parameterized black holes \cite{kz, Younsi:2016azx}; Einstein-Maxwell-Chern-Simon black hole \cite{cs}; Einstein-Maxwell-Dilaton-Axion black hole \cite{wei_observing_2013}; Kerr-Newman-Kasuya black hole \cite{knk}; Kerr-Perfect fluid dark matter black hole \cite{pfdm}; Kerr-de Sitter black hole \cite{kds}; Kerr-MOG black hole \cite{mog, wang_shadows_2019}; rotating regular black hole \cite{rr}; Kerr-Sen black hole \cite{Dastan:2016bfy}; charged rotating black hole in $f(R)$ gravity \cite{ks}, non-commutative black holes \cite{nc}, Tomimatsu-Sato spacetime \cite{Bambi:2010hf}, black hole surrounded by dark matter halo \cite{Jusufi:2019nrn,hou_black_2018, konoplya_shadow_2019}, shadow images of a rotating global monopole \cite{Haroon:2019new}, testing the rotational nature of the supermassive object M87 and possible signals for extra dimensions in the shadow of M87 \cite{vagnozzi1,vagnozzi2} etc, and also naked singularities \cite{ns, Bambi:2008jg}. Recently, there also are some interesting works about the curvature and topology of the geometry of the black hole shadows \cite{wei_intrinsic_2019, wei_curvature_2019}.

On the other hand, the Lorentz invariance is one of the fundamental principle of the general relativity (GR). However, when one considers the quantization of the gravity, such invariance could be violated at high energy regime. In this sense, the Lorentz symmetry can only be treated as a approximate symmetry, which emerges at low energies and violated at higher energies. With these thoughts, a lot of modified gravity theories have been proposed, such as the Horava-Lifshitz theories of quantum gravity (see refs. \cite{horava_quantum_2009, Zhu:2011yu, Wang:2017brl} for examples) and Einstein-\AE{}ther theory \cite{davi, d1, jacobson_einstein-aether_2008}.

Einstein-\AE ther theory is a generally covariant theory of gravitation which violates the Lorentz symmetry locally. The theory possesses the Riemannian metric along with a dynamical and timelike vector field (the \ae{}ther field) at each point of the spacetime manifold. The presence of the \ae{}ther field defines a preferred timelike direction that violates the Lorentzian symmetry unlike general relativity (GR) \cite{davi,d1}. The breaking of Lorentz symmetry might occur at the Planck or quantum gravity scales if the spacetime continuum is reduced to a discrete structure, thereby leading spacetime to be an emergent phenomenon. The Horava-Lifshitz theory is also a leading candidate for a quantum gravity and violates the Lorentz symmetry by suggesting an anisotropic scaling for time and space coordinates. Various implications of this kind of theories are explored in cosmological contexts such as cosmological perturbations \cite{ li, bat}, the effects of the parity violation in the gravitational waves \cite{Zhu:2013fja, Wang:2012fi}, and the inflation in the early universe \cite{ Zhu:2012zk, tedy, li, Zhu:2013fha}.
Moreover the astrophysical constraints on the coupling parameters of the theory are also studied with the data of the gravitational wave events GW170817 and GRB 170817A \cite{aw}.
The laws of black hole thermodynamics and the analysis of cosmic censorship conjecture involving the existence universal horizons during gravitational collapse have been investigated in \cite{af}.

Therefore, it is of great interesting to explore the properties of the black holes with the presence of the \ae{}ther field. Recently, two static, charged, and spherical symmetric black hole solutions have been found in the Einstein-\AE ther theory with two specific combination of the coupling constants \cite{ding_charged_2015}. {Other spherical symmetric black hole solution for a class of coupling constants has also been explored by using numerical calculation \cite{Eling:2006ec} and its analytical representation in the polynomial form has been used in the study of the quasi-normal modes in the Einstein-\AE{}ther theory. \cite{Konoplya:2006rv, Konoplya:2006ar}.} With these  spherical symmetric black hole solutions, one can apply the procedure in \cite{Wang:2012nv, barausse_slowly_2016} to generate the slow-rotating black hole solutions. These solutions, which contains the effects of the \ae{}ther field, can be used to study the effects of the \ae{}ther field in the black hole shadows and their constraints from the data of the first image of the M87 black hole. And this is exactly the purpose of the current paper.

The plan of our paper is as follows. In Sec. II, we present a a brief review of the charged and slowly rotating black hole solutions and their properties in the Einstein-\ae ther theory. The null geodesic equations and orbital equations of photons are given in Sec. III. Then in Sec. IV, we study the shadows of the two types of the black holes in the Einstein-\AE{}ther theory. The observational constraints on the \ae{}ther parameters by using the data of the first black hole image are presented in Sec. V. In Sec. VI we study the effect of \ae{}ther field on the deflection angle of light by two types of the Eisntein-\AE{}ther black holes. Moreover in Sec. VII we have studied the effect of \ae{}ther parameters on the time delay.  The summary and discussion for this paper is presented in Sec. VIII.

\section{Black hole solutions in Einstein-\AE ther theory}
\renewcommand{\theequation}{2.\arabic{equation}} \setcounter{equation}{0}

In this section, we present a brief review of the black hole solutions in the Einstein-\ae ther theory.

\subsection{Field Equations of the Einstein-\AE{}ther theory}

In Einstein-\ae ther theory, in addition to the spacetime metric tensor field $g_{\mu\nu}$, it involves a dynamical, unit timelike \ae ther field $u^{\alpha}$ (it is also called \ae ther four velocity) \cite{jacobson_gravity_2001, foster_radiation_2006, garfinkle_numerical_2007, jacobson_einstein-aether_2008}. Like the metric, and unlike other classical fields, the \ae ther field $u^{\alpha}$ cannot vanish anywhere, so it breaks local Lorentz symmetry. With this unit and timelike vector, the general action of the Einstein-\ae ther theory is given by \cite{jacobson_einstein-aether_2008}
\bqn
S_{\ae} = \frac{1}{16 \pi G_{\ae}} \int d^4 x \sqrt{-g} \Big(R+ \mathcal{L}_{\ae}\Big),
\eqn
where $g$ is the determinant of the four dimensional metric $g_{\mu\nu}$ of the space-time with the signatures $(-, +, +, +)$, $R$ is the Ricci scalar, $G_{\ae}$ is the the aether gravitational constant, and the Lagrangiean of the \ae ther field $\mathcal{L}_{\ae}$ is given by
\bqn
\mathcal{L}_{\ae} \equiv -M^{\alpha \beta}_{\;\; \;\; \mu\nu} (D_\alpha u^\mu) (D_{\beta} u^\nu) + \lambda (g_{\mu \nu} u^\mu u^\nu +1).\nb\\
\eqn
Here $D_{\alpha}$ denotes the covariant derivative with respect to $g_{\mu\nu}$, $\lambda$ is a Lagrangian multiplier, which guarantees that the aether four-velocity $u^{\alpha}$ is always timelike, and $M^{\alpha \beta}_{\;\; \;\; \mu\nu}$ is defined as \footnote{The parameters $(c_1, c_2, c_3, c_4)$ used in this paper are related to parameters $(c_\theta, c_\sigma, c_\omega, c_a)$ by the relations $c_{\theta} = c_1+3c_2+c_3$, $c_{\sigma } = c_1+c_3 = c_{13}$, $c_\omega = c_1- c_3$, $c_a = c_1+c_4 = c_{14}$.}
\bqn
M^{\alpha \beta}_{\;\; \;\; \mu\nu} \equiv c_1 g^{\alpha \beta} g_{\mu\nu} + c_2 \delta^{\alpha}_{\mu} \delta ^{\beta}_{\nu} + c_3  \delta^{\alpha}_{\nu} \delta ^{\beta}_{\mu} - c_4 u^{\alpha} u^{\beta} g_{\mu\nu}. \nb\\
\eqn
The four coupling constants $c_i$ ' s are all dimensionless, and $G_{\ae}$ is related to the Newtonian constant $G_N$ via the relation \cite{carroll_lorentz-violating_2004},
\bqn
G_{\ae} = \frac{G_N}{1-\frac{1}{2}c_{14}}
\eqn
with $c_{14} \equiv c_1+c_4$. In order to discuss the black hole solution with electric change, we also add a source-free Maxwell Lagrangian $\mathcal{L}_{M}$ to the theory, then the total action of the theory becomes,
\bqn
S_{\ae, M} =  S_{\ae} + \int d^4x \sqrt{-g } \mathcal{L}_M,
\eqn
where
\bqn
\mathcal{L}_{M} = - \frac{1}{16 \pi G_{\ae}} F^{\mu\nu}F_{\mu\nu},\\
F_{\mu\nu} = D_\mu A_\nu - D_\nu A_\mu,
\eqn
where $A_\mu$ is the electromagnetic potential four-vector. {It is worth noting that the electromagnetic field $A_\mu$ is minimally coupled to the gravity and the \ae{}ther field $u_\mu$.}

The variations of the total action with respect to $g_{\mu\nu}$, $u^{\alpha}$, $\lambda$, and $A^a$ yield, respectively, the field equations,
\bqn
E^{\mu\nu} =0, \lb{einstein_equation}\\
{\AE}_{\alpha} =0, \lb{aether_equation}\\
g_{\mu\nu} u^{\mu} u^{\nu} =-1 \lb{lambda_equation},\\
D^\mu F_{\mu\nu}=0.
\eqn
where
\bqn
E^{\mu\nu} \equiv R^{\mu\nu} - \frac{1}{2} g^{\mu\nu} R - 8 \pi G_{\ae} T_{\ae}^{\mu\nu},\\
\AE_{\alpha} \equiv D_{\mu} J^{\mu}_{\;\;\;\alpha} + c_4 a_{\mu} D_{\alpha}u^{\mu} + \lambda u_{\alpha},
\eqn
with
\bqn
T^{\ae}_{\alpha\beta} &\equiv& D_{\mu}\Big[J^{\mu}_{\;\;\;(\alpha}u_{\beta)} + J_{(\alpha\beta)}u^{\mu}-u_{(\beta}J_{\alpha)}^{\;\;\;\mu}\Big]\nb\\
&& + c_1\Big[\left(D_{\alpha}u_{\mu}\right)\left(D_{\beta}u^{\mu}\right) - \left(D_{\mu}u_{\alpha}\right)\left(D^{\mu}u_{\beta}\right)\Big]\nb\\
&& + c_4 a_{\alpha}a_{\beta}    + \lambda  u_{\alpha}u_{\beta} - \frac{1}{2}  g_{\alpha\beta} J^{\delta}_{\;\;\sigma} D_{\delta}u^{\sigma},\nb\\
J^{\alpha}_{\;\;\;\mu} &\equiv& M^{\alpha\beta}_{~~~~\mu\nu}D_{\beta}u^{\nu},\nb\\
a^{\mu} &\equiv& u^{\alpha}D_{\alpha}u^{\mu}.
\eqn
From Eqs.(\ref{aether_equation}) and (\ref{lambda_equation}),  we find that
\bqn
\lb{2.7}
\lambda = u_{\beta}D_{\alpha}J^{\alpha\beta} + c_4 a^2,
\eqn
where $a^{2}\equiv a_{\lambda}a^{\lambda}$.

\subsection{Static and Charged Spherically Symmetric Einstein-\ae ther Black Holes}

The general form for a static spherically symmetric metric for Einstein-\ae ther black hole spacetimes can be written in the Eddington-Finklestein coordinate system as
\bqn \lb{metric_EF}
ds^2 = -e(r) dv^2 +2 f(r) dv dr + r^2 (d\theta^2+\sin^2\theta d\phi^2),\nb\\
\eqn
with the corresponding Killing vector $\chi^a$ and the \ae ther vector field $u^{a}$ being given by
\bqn
\chi^a=(1,0,0,0),\;\;\; u^{a}(r) = (\alpha(r), \beta(r), 0, 0 ),
\eqn
where $e(r)$, $f(r)$, $\alpha(r)$, and $\beta(r)$ are functions of $r$ only.
The boundary conditions on the metric components are such that the solution is asymptotically flat, while those for the \ae{}ther components are such that
\bqn
\lim_{r \to +\infty} u^a = (1,0,0,0).
\eqn

As shown in \cite{ding_charged_2015}, there exist two types of exact static and charged spherically symmetric black hole solutions in Einstein-\ae ther theory. The first solution corresponds to the special choice of coupling constants $c_{14}=0$ and $c_{123} \neq 0$ where $c_{123} \equiv c_1+c_2+c_3$, while the second solution corresponds to $c_{123}=0$. 

\subsubsection{$c_{14}=0$ and $c_{123} \neq 0$}

For the first solution, we have \cite{ding_charged_2015}
\bqn
e(r) &=& 1- \frac{2M}{r} +\frac{Q}{r^2} \nb\\
&&- \frac{27 c_{13}}{256(1-c_{13})} \left(\frac{2M}{r} \right)^4,\lb{e14}\\
f(r) &=& 1,\\
\alpha(r) &=& \Bigg[ \frac{1}{\sqrt{1-c_{13}}} \frac{3\sqrt{3}}{16} \left(\frac{2M}{r}\right)^2 \nb\\
&&~~~  + \sqrt{1-\frac{2M}{r} + \frac{27}{256} \left(\frac{2M}{r}\right)^4}\Bigg]^{-1},\\
\beta(r) &=& - \frac{1}{\sqrt{1-c_{13}}} \frac{3 \sqrt{3}}{16} \left(\frac{2M}{r}\right)^2.
\eqn
Here $M$ and $Q$ are the mass and the electric charge of the black hole spacetime respectively. It is obvious that when $c_{13}=0$, the above solution reduces to the Reissner-Nordstr\"{o}m black hole.

\subsubsection{$c_{123}=0$}

For the second solution, we have \cite{ding_charged_2015}
\bqn
e(r) &=& 1- \frac{2M}{r} + \frac{Q}{(1-c_{13})r^2} \nb\\
&&-\frac{2c_{13} - c_{14}}{8(1-c_{13})} \left(\frac{2M}{r} \right)^2, \lb{e123}\\
f(r) &=& 1,\\
\alpha(r) &=&  \frac{1}{1+\frac{1}{2} \left[\sqrt{\frac{2-c_{14}}{2(1-c_{13})}} -1 \right]\frac{2M}{r}},\\
\beta(r) &=& - \frac{1}{2} \sqrt{\frac{2-c_{14}}{2(1-c_{13})}} \frac{2M}{r} .
\eqn
In this case, when $c_{13}=0=c_{14}$, it also reduces to the Reissner-Nordstr\"{o}m black hole.

For both solutions, it is convenient to write the metric (\ref{metric_EF}) in the the Eddington-Finklestein coordinate system in the form of the usual $(t, \; r,\; \theta, \; \phi)$ coordinates. This can be achieved by using the coordinate transformation
\bqn
dt = dv - \frac{dr }{e(r)}, \;\; dr=dr.
\eqn
Then the metric of the background spacetime turns into the form
\bqn\lb{metric}
ds^2 = - e(r)dt^2+\frac{dr^2}{e(r)} +  r^2 (d\theta^2+\sin^2\theta d\phi^2).
\eqn
In this metric, the \ae{}ther field reads
\bqn
u^a= \left(\alpha(r) - \frac{\beta(r)}{e(r)}, \beta(r), 0, 0\right).
\eqn


\subsection{Slowly rotating black holes}

The rotating black hole in the slow rotation limit in general can be described by the the well-known Hartle-Thorne metric \cite{hartle_slowly_1968}
\bqn
ds^2 &=& -e(r) dt^2 + \frac{B(r) dr^2}{e(r)} +  r^2 (d\theta^2+\sin^2\theta d\phi^2) \nb\\
&& -\epsilon r^2 \Omega(r, \theta) dt d\phi + \mathcal{O}(\epsilon^2),
\eqn
where $e(r)$ represents the ``seed" static, spherically-symmetric solutions when $\Omega(r, \theta) =0$, $\epsilon$ denotes a small perturbative rotation parameter. For discussions of the black holes in this paper, we consider $B(r)=1$ \cite{Wang:2012nv, barausse_slowly_2016}. The \ae{}ther configuration in the slow-rotation limit is described by \cite{barausse_slowly_2016}
\bqn
u_a dx^a &=&[\beta(r)-e(r) \alpha(r)]dt +\frac{\beta(r)}{e(r)} dr \nb\\
&&+ \epsilon  [\beta(r)-e(r) \alpha(r)]\lambda(r, \theta) \sin^2\theta d \phi + \mathcal{O}(\epsilon^2),\nb\\
\eqn
where $\lambda(r, \theta)$ is related to the \ae{}ther's angular momentum per unit energy by $u_\phi/u_t = \lambda(r, \theta) \sin^2\theta$.

For asymptotically flat boundary condition, as shown in \cite{barausse_slowly_2016}, $\Omega(r,\theta)$ and $\lambda(r, \theta)$ have to be $\theta$-independent, namely $\Omega(r, \theta)=\Omega(r)$ and $\lambda(r, \theta) = \lambda(r)$. Then the \ae{}ther's angular velocity can be written as
\bqn
\psi(r)=\frac{u^\phi}{u^t}= \frac{1}{2}\Omega(r)- \frac{\lambda(r)}{r^2}.
\eqn
The slowly rotating black holes in the Einstein-\AE ther theory have been obtained in \cite{Wang:2012nv, barausse_slowly_2016} and have been discussed in several papers that related to Horava-Lifshitz gravity \cite{Wang:2012nv, Wang:2012nv}. While in these mentioned papers only neutral black holes have been considered, in this subsection, we present the metric of the charged slowly rotating black holes by directly applying the forms in \cite{barausse_slowly_2016} for $c_{14} =0$ but $c_{123} \neq 0$ and $c_{123}=0$ respectively.

\subsubsection{$c_{14} =0$ and $c_{123} \neq 0$}
For this case, there exists slowly rotating black hole solution in the Einstein-\AE ther theory with a spherically symmetric (hypersurface-orthogonal) \ae ther field configuration, which leads to
\bqn
\Omega(r)= \frac{4 J}{r^3}\;\; {\rm and}\;\; \lambda(r)=0.
\eqn
The metric now reads
\bqn\lb{slow1}
ds^2 &=& -e(r) dt^2 + \frac{dr^2}{e(r)} +  r^2 (d\theta^2+\sin^2\theta d\phi^2)j \nb\\
&& -\frac{4 M}{r} a \sin^2 \theta dt d\phi + \mathcal{O}(\epsilon^2).
\eqn
Note that $e(r)$ is given by Eq.~(\ref{e123}).

\subsubsection{$c_{123}=0$}

For this case, there is no closed form for the expression of $\Omega(r)$ and $\lambda(r)$ except in the limit $c_\omega = c_1-c_3 \to \infty$. In \cite{barausse_slowly_2016}, the asymptotic forms for the derivatives $\Omega'(r)$ and $\lambda'(r)$ has been obtained by expanding them for large $r$ and the corresponding integration constants can be determined by using numerical calculation. In the limit $c_\omega \to \infty$, the frame dragging potential $\Omega(r)$ has the form \cite{barausse_slowly_2016}
\bqn
\Omega(r)=\frac{4 J}{4}.
\eqn
Then the metric in this case takes the same form as (\ref{slow1}) but with $e(r)$ given by Eq.~(\ref{e14}). As pointed out in \cite{Wang:2012nv},
the Einstein-\ae{}ther theory in the limit $c_\omega \to \infty$ coincides to the non-projectable Horava-Lifshitz theory of gravity in the infrared limit, thus the solutions in the Einstein-\ae{}ther theory with $c_\omega \to \infty$ are also solutions of the Horava-Lifshitz theory of gravity.

{
\subsection{Numerical black hole solution}
Except the black hole solutions in the analytical form, the numerical black hole solution has also been explored in the Einstein-\AE{}ther theory for a class of coupling constants \cite{Eling:2006ec}. For spherically symmetric solution, $c_4$ can be absorbed into other coupling constants. In \cite{Eling:2006ec},  the so-called non-reduced Einstein-\AE{}ther theory is considered, for which $c_3=0$, and then one can use the field redefinition that fixes the coefficient $c_2$ \cite{Eling:2006ec, Konoplya:2006rv, Konoplya:2006ar} ,
\bqn
c_2= - \frac{-2 c_1^3}{2-4 c_1 +3 c_1^3},
\eqn
so that $c_1$ is the only free parameter in the numerical black hole solution in \cite{Eling:2006ec}. With these set-up, the metric for a spherically symmetric static black hole can still be described by (\ref{metric_EF}), in which the functions $e(r)$, $f(r)$, $\alpha(r)$, and $\beta(r)$ are determined by numerical calculations. An approximate form of $e(r)$ and $f(r)$ are given in the polynomial form in \cite{Konoplya:2006rv, Konoplya:2006ar} as
\bqn
e(r) = \frac{\sum_{i=0}^{N_e} a_{i}^{(e)} r^i} {1+\sum_{i=0}^{N_e} b_{i}^{(e)} r^i},\\ 
f(r)=  \frac{\sum_{i=0}^{N_f} a_{i}^{(f)} r^i} {1+\sum_{i=0}^{N_f} b_{i}^{(f)} r^i},
\eqn
where the coefficients $(a_i^{(e)}, b_i^{(e)}; a_i^{(f)}, b_i^{(f)})$ can be determined by fitting the above analytical form with numerical solutions, which is out of the scope of the current paper. The general features of the this numerical black hole has been discussed in \cite{Eling:2006ec, Konoplya:2006rv, Konoplya:2006ar}, and it has been shown that the \ae{}ther parameter $c_1$ trends to decrease the radius of the horizon.
}

%

\section{Null Geodesics and photon orbits}
\renewcommand{\theequation}{3.\arabic{equation}} \setcounter{equation}{0}

In this section, we analyze the evolution of the photon around the black holes in the Einstein-\ae{}ther theory.  {Here we ignore the interaction between the photon and the \ae{}ther field, in this way, the presence of the \ae{}ther field only affects the background black hole spacetime and the photon follows the null geodesics in such given black hole spacetime.} Therefore for the purpose of studying the photon trajectories, one needs to first study its geodesics structure, which is described by the Hamilton-Jacobi equation,
\bqn\lb{HJ}
\frac{\partial S}{\partial \lambda} = - \frac{1}{2}g^{\mu\nu} \frac{\partial S}{\partial x^\mu} \frac{\partial S}{\partial x^\nu},
\eqn
where $\lambda$ is the affine parameter of the null geodesic and $S$ denotes the Jacobi action of the photon. The Jacobi action $S$ can be separated in the following form,
\bqn\lb{Jacobi_action}
S = \frac{1}{2} m^2 \lambda - E t + L \phi + S_{r}(r) + S_\theta (\theta),
\eqn
where $m$ denotes the mass of the particle moving in the black hole spacetime and for photon one has $m=0$. $E$ and $L$ represent the energy and angular momentum of the photon in the direction of the rotation axis respectively. The two functions $S_r(r)$ and $S_\theta(\theta)$ depend only on $r$ and $\theta$ respectively.

Now we can substitute the Jacobi action (\ref{Jacobi_action}) into the Hamilton-Jacobi equation (\ref{HJ}) we obtain
\bqn
-m^2 r^2 &=&  r^2 e(r) \left(\frac{dS_r}{dr}\right)^2 - \frac{E^2 r^2}{e(r)} + \frac{4 E L M a}{r e(r)}  \nb\\
&&+ L^2 \csc^2 \theta + \left(\frac{dS_\theta}{d\theta}\right)^2 + \mathcal{O}(a^2).
\eqn
Note that in derivation of the above equation, we have used the metric in (\ref{slow1}). It is immediately observed that this equation is separable and for the photon ($m=0$) we have
\bqn
L^2 \csc^2 \theta \cos^2\theta+ \left(\frac{dS_\theta}{d\theta}\right)^2 = \mathcal{K},\\
 r^2 e(r) \left(\frac{dS_r}{dr}\right)^2 - \frac{E^2 r^2}{e(r)} + \frac{4 E L M a}{r e(r)} = - \mathcal{K}-L^2,
\eqn
where $\mathcal{K}$ is the separation constant. Then the solution of $S_\theta(r)$ and $S_r(r)$ can be written as
\bqn
S_{\theta} (\theta) = \int^\theta \sqrt{\Theta(\theta)} d\theta, \\
S_r(r) = \int^r \frac{\sqrt{R(r)}}{r^2 e(r)} dr,
\eqn
where one introduces
\bqn
R(r) &=&  E^2r^4 - (\mathcal{K}+L^2) r^2 e(r) - 4 M a EL r, \\
\Theta(\theta) &=& \mathcal{K} - L^2 \csc^2 \theta \cos^2\theta.
\eqn
Then variation of the Jacobi action gives the following four equations of motion for the evolution of the photon,
\bqn
\frac{dt}{d\lambda} &=& \frac{1}{e(r)} \left(E - \frac{2 M a L}{r^3}\right), \\
\frac{dr}{d\lambda} &=& \frac{\sqrt{R(r)}}{r^2}, \\
\frac{d\theta}{d\lambda} &=& \frac{\sqrt{\Theta(\theta)}}{r^2}, \\
\frac{d \phi}{d \lambda} &=& \frac{L \csc^2\theta}{r^2} - \frac{2 M E a}{r^3 e(r)}.
\eqn

\begin{figure*}
\includegraphics[width=8.1cm]{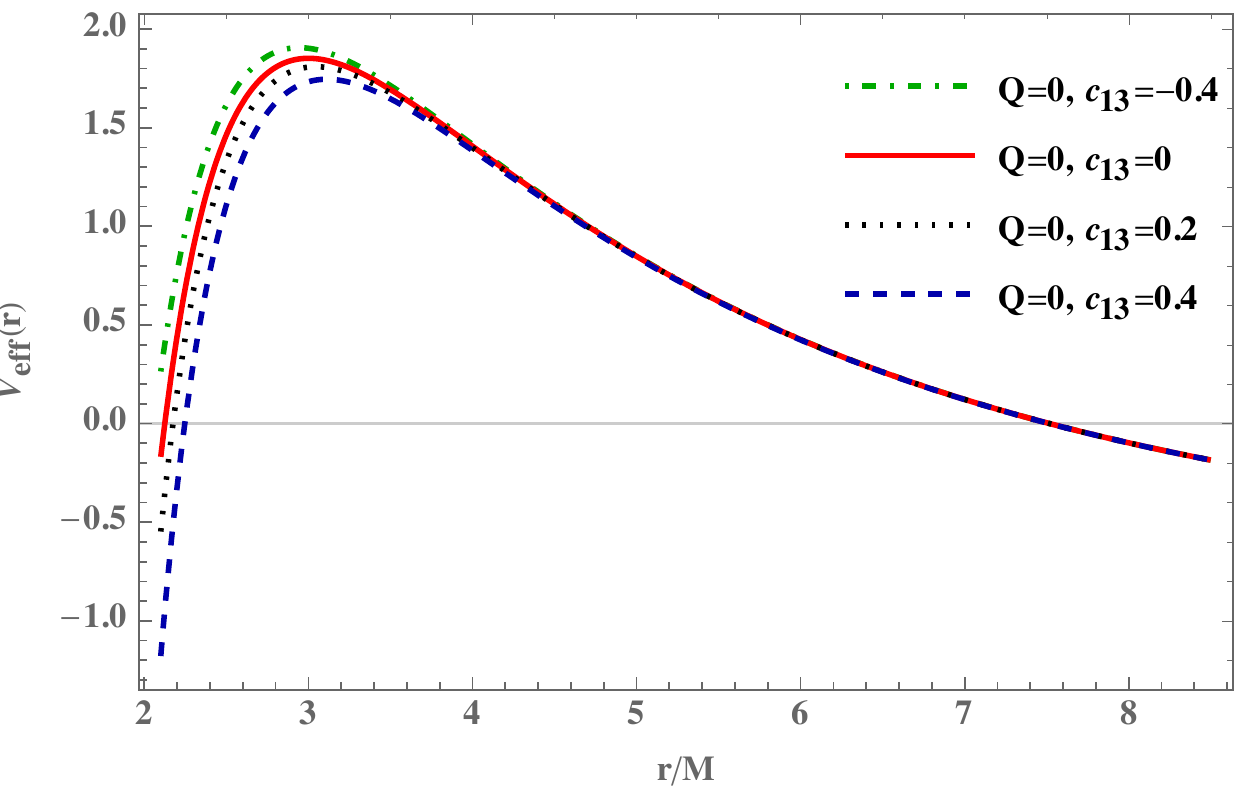}
\includegraphics[width=8.1cm]{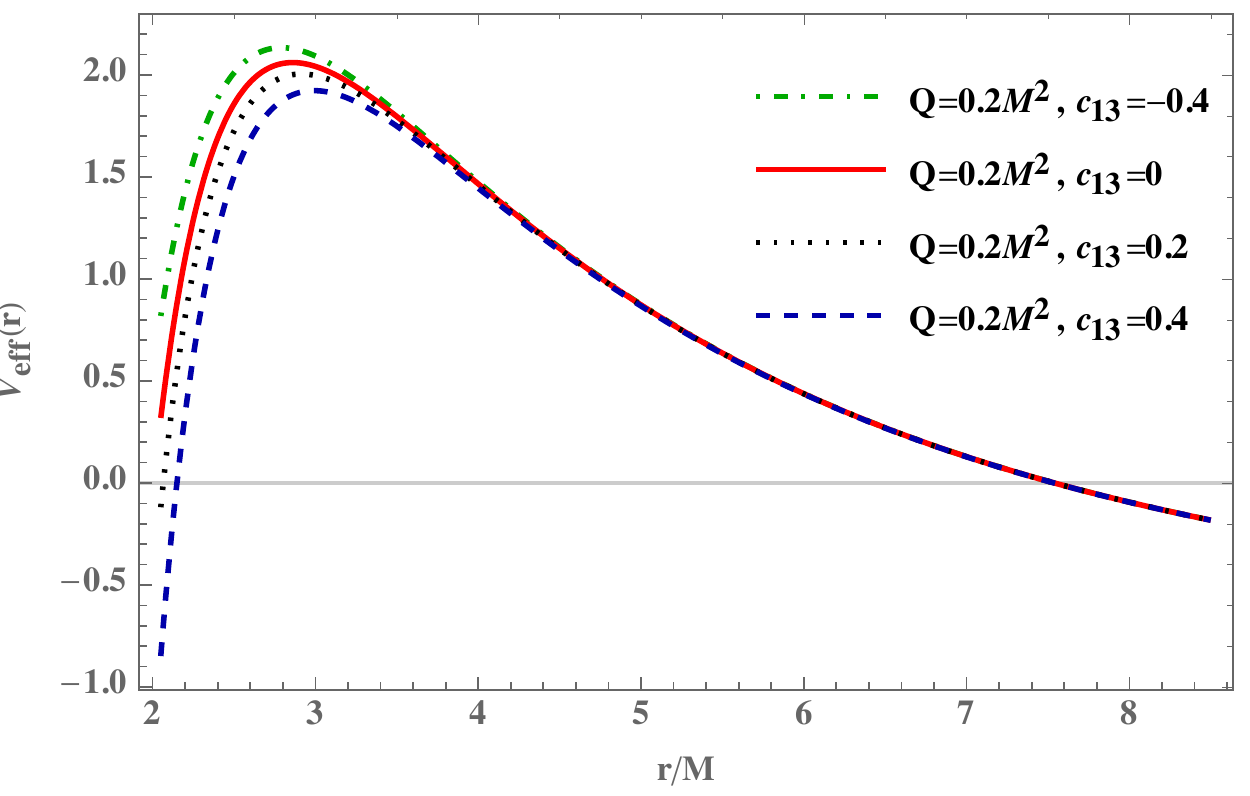}
\caption{The radial dependence of the effective potential of the massless particles in the first type Einstein-\AE{}ther black hole for the different values of the $c_{13}$. Left panel: for static and neutral black hole. Right panel: for static and charged black hole.} \label{Veff1}
\end{figure*}

To study the trajectories of photons, it is convenient to write the radial geodesics in terms of the effective potential $V_{\rm eff}(r)$ as
\bqn
 \left(\frac{dr}{d\lambda}\right)^2 + V_{\rm eff} (r)= 0
\eqn
with
\bqn
V_{\rm eff}(r) &=& - r^{-4}R(r)/E^2 \nb\\
&=& -1 + \frac{e(r)}{r^2} (\xi^2 +\eta) + \frac{4 M a \eta}{r^3},
\eqn
where we define
\bqn
\xi = \frac{L}{E},\;\;\eta = \frac{\mathcal{K}}{E^2}.
\eqn
The motion of the photon can be determined by these two impact parameters.
In Fig.~\ref{Veff1}, left and right panels, we have plotted the effective potential obeyed by the photons in the background of the first type static neutral and charged \AE{}ther black holes (with $c_{14}=0$ but $c_{123} \neq 0$). It is observed that the $V_{\rm eff}$ admits a unique maximum which depicts the presence of an unstable circular orbit around $r/M\simeq 3$. Thus photons can leave the circular orbit if perturbed by external gravitational forces or by interacting with other particles. These photons in turn constitute a photon sphere which is observable as a black hole shadow in the observers's frame. The photons escaping from the nearest orbit around black hole provide significant information about the horizon geometry, spin and size as well. For the neutral BH ($Q=0$), by increasing the value of $c_{13}$ parameter, the peak of $V_{\rm eff}$ gets shorter, while similar effects are observed if the black hole is charged. In the asymptotic regime, the $V_{\rm eff}$ approaches the limit $-1$ which represents a constant energy of the photon and its motion remains stable at infinity.

\begin{figure*}
\includegraphics[width=8.1cm]{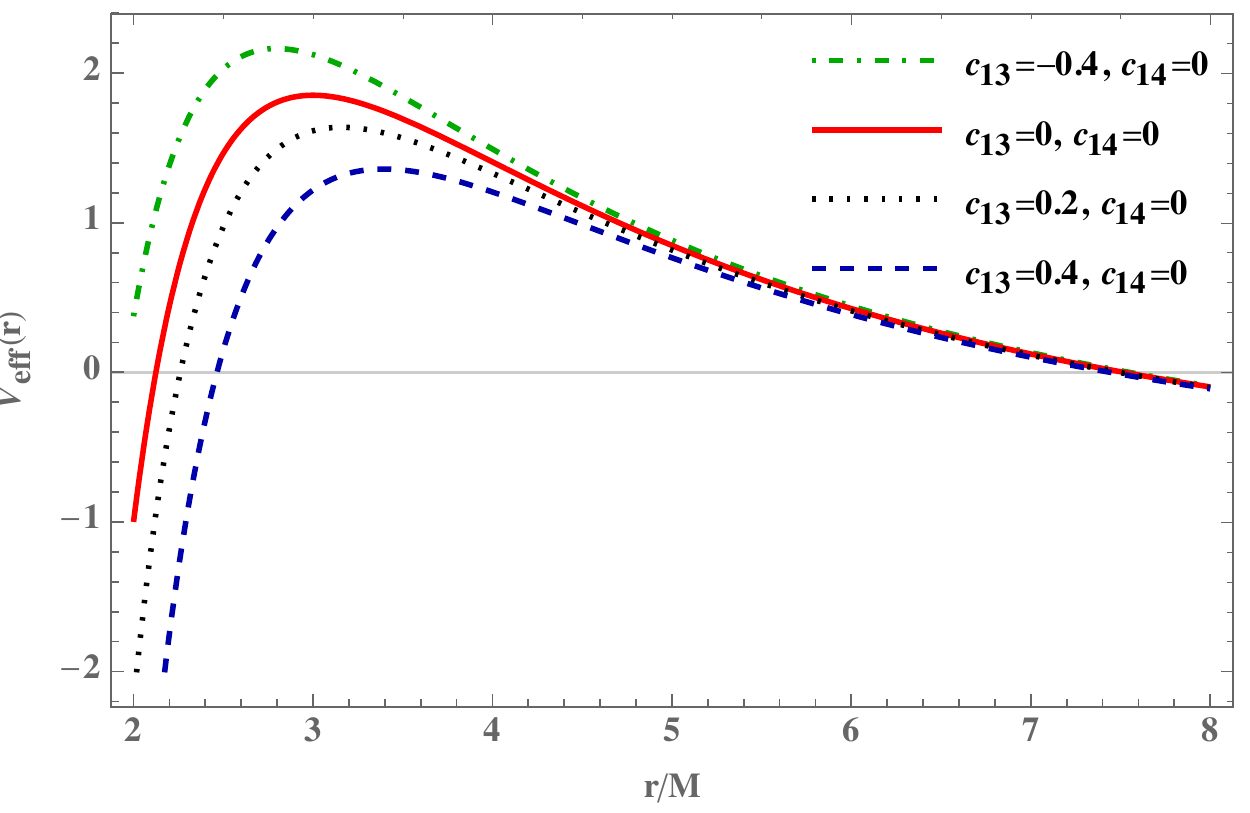}
\includegraphics[width=8.1cm]{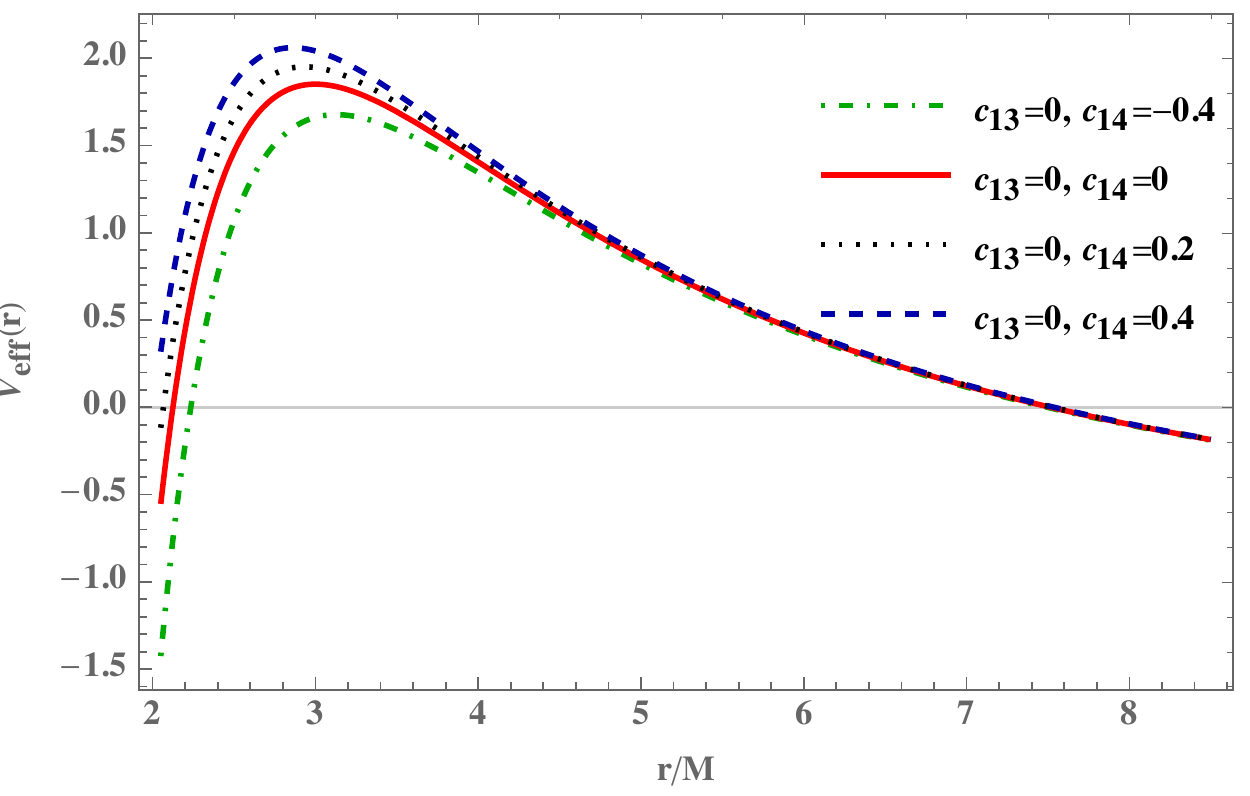}
\caption{The radial dependence of the effective potential of the massless particles in the second type Einstein-\AE{}ther black hole for the different values of the $c_{13}$. Left panel: for static and neutral black hole with $c_{14}=0$. Right panel: for static and neutral black hole with $c_{13}=0$.} \label{Veff2}
\end{figure*}

In Fig.~\ref{Veff2}, left and right panels, we have plotted the effective potential energy of the photons around the second type neutral and charged Einstein-\AE{}ther black hole (with $c_{123}=0$), for different values of $c_{13}$ and $c_{14}$ parameters. Note that by fixing one parameter and varying the other, the curves interchange their position. For instance in left panel, by taking $c_{14}=0$ and increasing $c_{13}$ leads the peak/maximum of the $V_{\rm eff}$ to decrease. The opposite effect occurs in the right panel when $c_{13}=0$ and $c_{14}$ is varied, causing the peak/maximum to increase. The values of these parameters as used in the figures are consistent with the astrophysical data. Further, the position of the maximum of $V_{\rm eff}$ does not drastically change until the parameters take very small or large values.

To determine the geometric sharp of the shadow of the black hole, we need to find the critical circular orbit for the photon, which can be derived from the unstable condition
\bqn\lb{condition}
R(r)=0,\;\; \frac{dR(r)}{dr} =0 ,\;\;\; \frac{d^2 R(r)}{dr^2} >0.
\eqn
Then the geometric sharp of the shadow can be determined by the allowed values of $\xi$ and $\eta$ that fulfill these conditions. In general, the sharp of the shadow depends on whether the rotation of the black hole is considered.

For spherical symmetric black holes, the shadow has spherical symmetry as well and is described by the photon sphere. In this case, the above conditions (\ref{condition}) can be simplified into,
\bqn\lb{PSradius}
2 - \frac{re'(r)}{e(r)}=0.
\eqn
The solution of this equation determines the radius $r_{\rm ps}$ of the photon sphere and one can write $\xi^2+\eta$ in the following form
\bqn
\xi^2 +\eta = \frac{r^2_{\rm ps}}{ e(r_{\rm ps})}.
\eqn
For slowly rotating black holes, solving the conditions (\ref{condition}), one immediately find that for the spherical orbits motion of photon, the two parameters $\xi$ and $\eta$ has the form
\bqn
\xi^2(r) &=& \frac{12 M r^2 a + 2 r^3 e(r) - r^4 e'(r)}{4 M a [e(r)+r e'(r)]}, \\
\eta(r) &=& \frac{r^3 [r e'(r) - 2\,e(r)]}{4 M a [e(r)+r e'(r)]} .
\eqn

{It is interesting to note that, for the spherically symmetric numerical black hole solution \cite{Eling:2006ec} and its representation in the polynomial form \cite{Konoplya:2006rv, Konoplya:2006ar}, one can use (\ref{PSradius}) to determine $r_{\rm ps}$ numerically. However, this strongly relies on the numerical calculation of the field equations in the Einstein-\AE{}ther theory, which is out of the scope of the current paper. Therefore, we are going to concentrate only on the two types of the analytical solutions in the next section.}

\section{Shadows of black holes}
\renewcommand{\theequation}{4.\arabic{equation}} \setcounter{equation}{0}

In general, the photon emitted by a light source will get deflected when it passes by a black hole because of the gravitational lensing effects. Some of the photon can reaches the distant observer after being deflected by the black hole, and some of them directly fall into the black hole. The photons that can not escape the black hole forms the shadow of the black hole in the observer's sky. To describe the shadow as seen by the distant observer, it is convenient to define the two celestial coordinates $X$ and $Y$ as,
\bqn
X &=& \lim_{r_* \to \infty} \left(- r_*^2 \sin \theta_0 \frac{d\phi}{dr}\right), \\
Y &=& \lim_{r_* \to \infty} r_*^2 \frac{d\theta}{dr},
\eqn
where $r_*$ denotes the distance between the observer and the black hole and $\theta_0$ represents the inclination angle between the line of sight of the observer and the rotational axis of the black hole. Since
\bqn
\frac{d\phi}{dr} = \frac{d\phi/d\lambda}{dr/d\lambda}, \;\; \frac{d\theta}{dr} = \frac{d\theta/d\lambda}{dr/d\lambda}, \nb
\eqn
then employing the geodesics equations we find
\bqn
X = - \xi(r_{\rm ps}) \csc\theta_0, \\
Y =  \sqrt{\eta (r_{\rm ps}) - \xi^2(r_{\rm ps}) \cot^2\theta_0}.
\eqn
It is easy to observe that the two celestial coordinates satisfy
\bqn
X^2+Y^2 = \xi^2(r_{\rm ps}) +\eta_{\rm ps}(r_{\rm ps}).
\eqn

This equation represents the shadow of the Einstein-\AE ther black holes in the slow rotation limit (only accurate up to $\mathcal{O}(a)$), which indicates the geometric shape of the shadow is still circular. The small rotation parameter $a=J/M \ll1$ can only affect the radius and the shape of the circular shadow if we consider its effects beyond the leading order. In this case one can still use (\ref{PSradius}) to determine the size of the shadow of the slow-rotating black holes.

\begin{figure*}
\includegraphics[width=8.1cm]{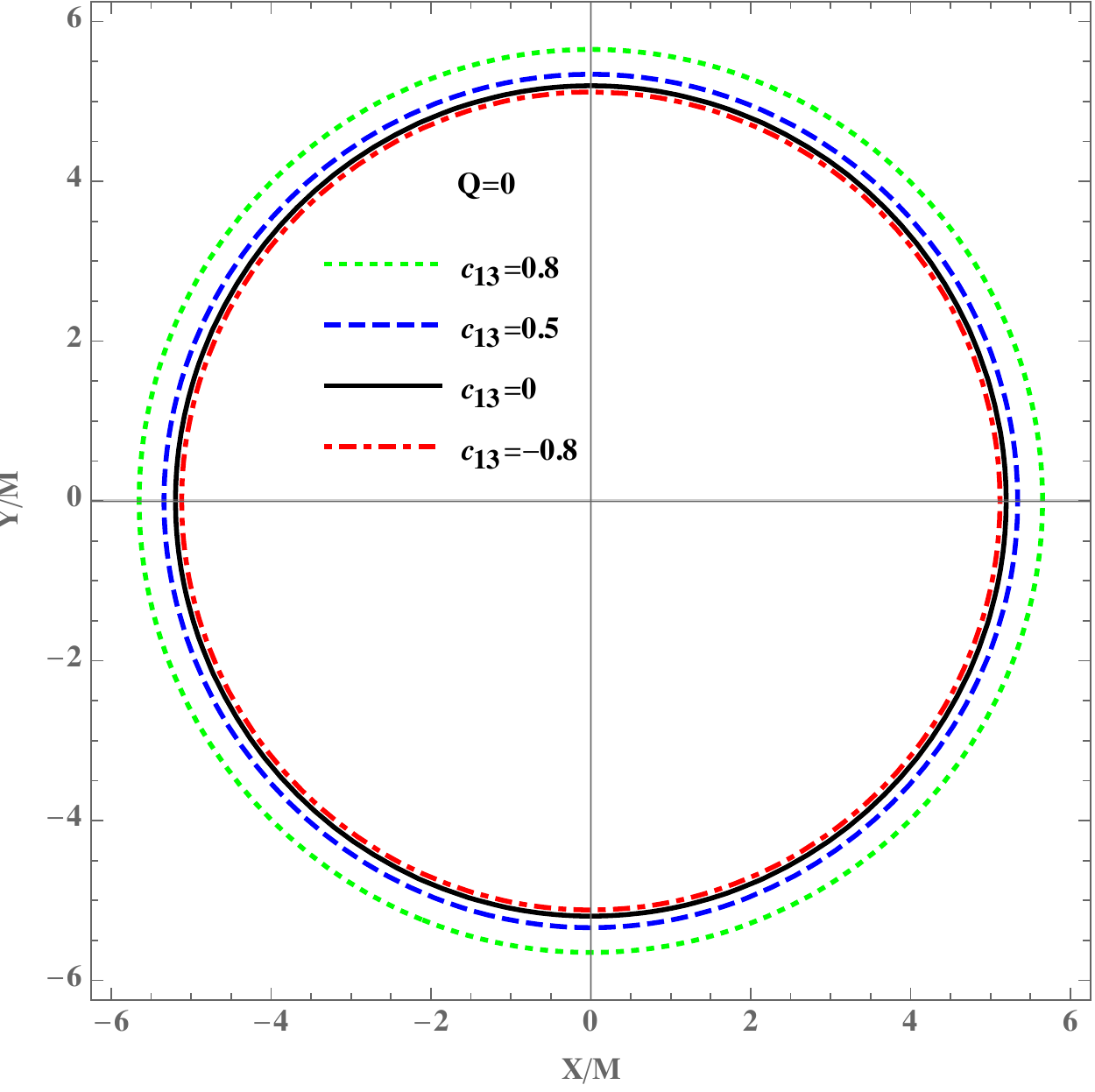}
\includegraphics[width=8.1cm]{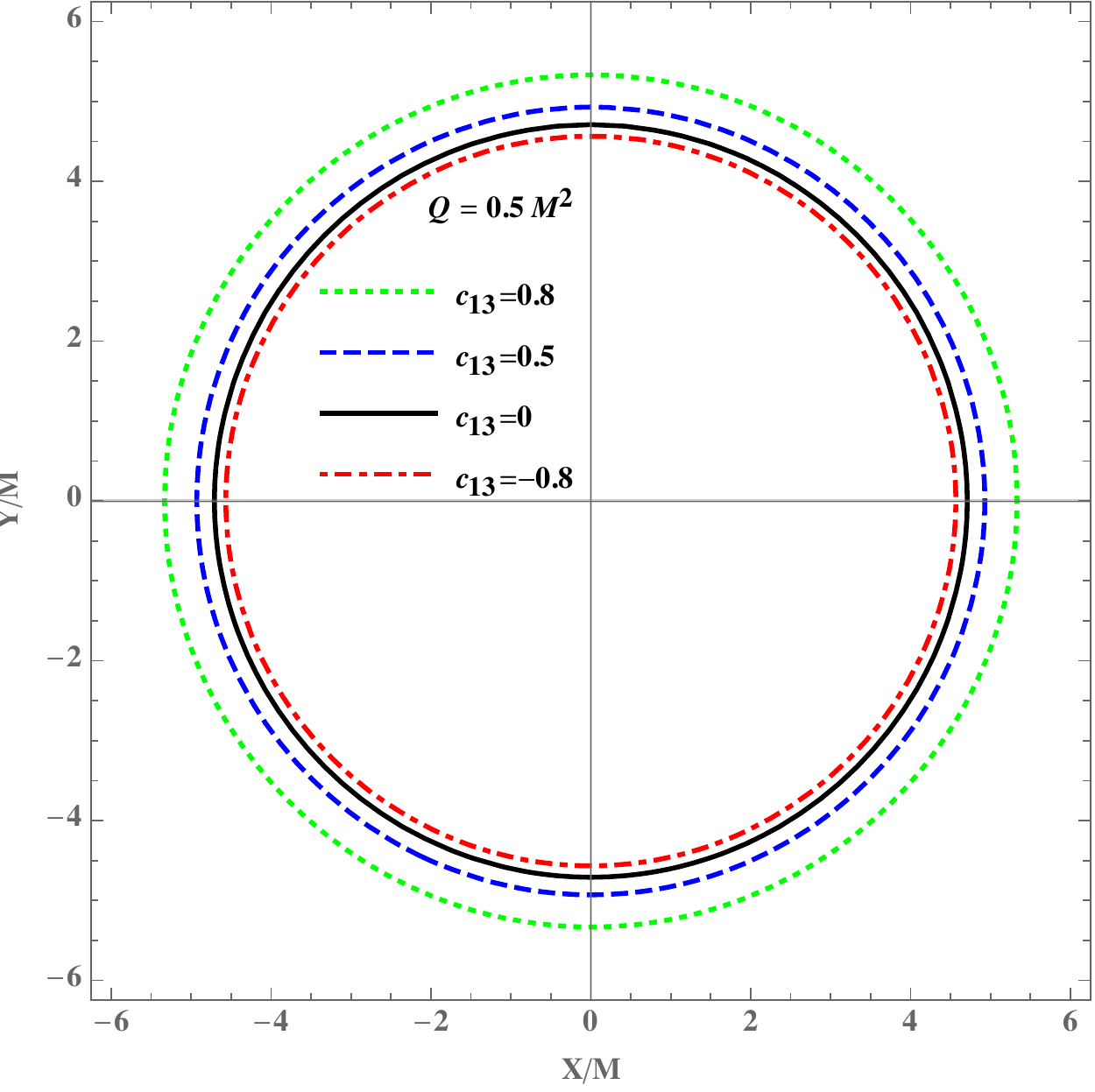}
\caption{Shadow region of the first spherical symmetric Einstein-\ae{}ther black hole ($c_{14}=0$ but $c_{123} \neq 0$). Left panel: the neutral case with $Q=0$. Right panel: changed case with $Q=0.5M^2$. } \label{A}
\end{figure*}

\begin{figure*}
\includegraphics[width=8.1cm]{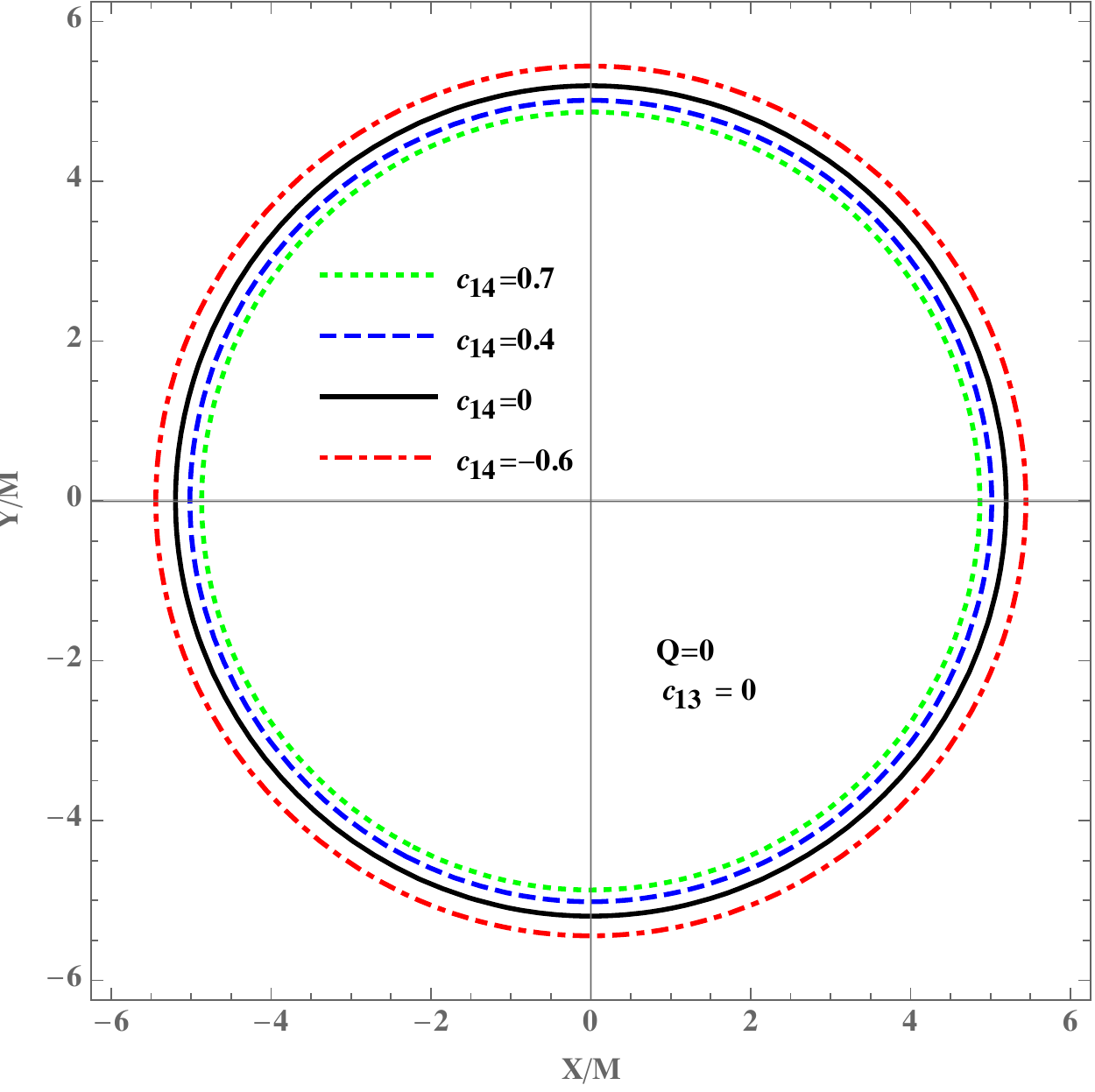}
\includegraphics[width=8.1cm]{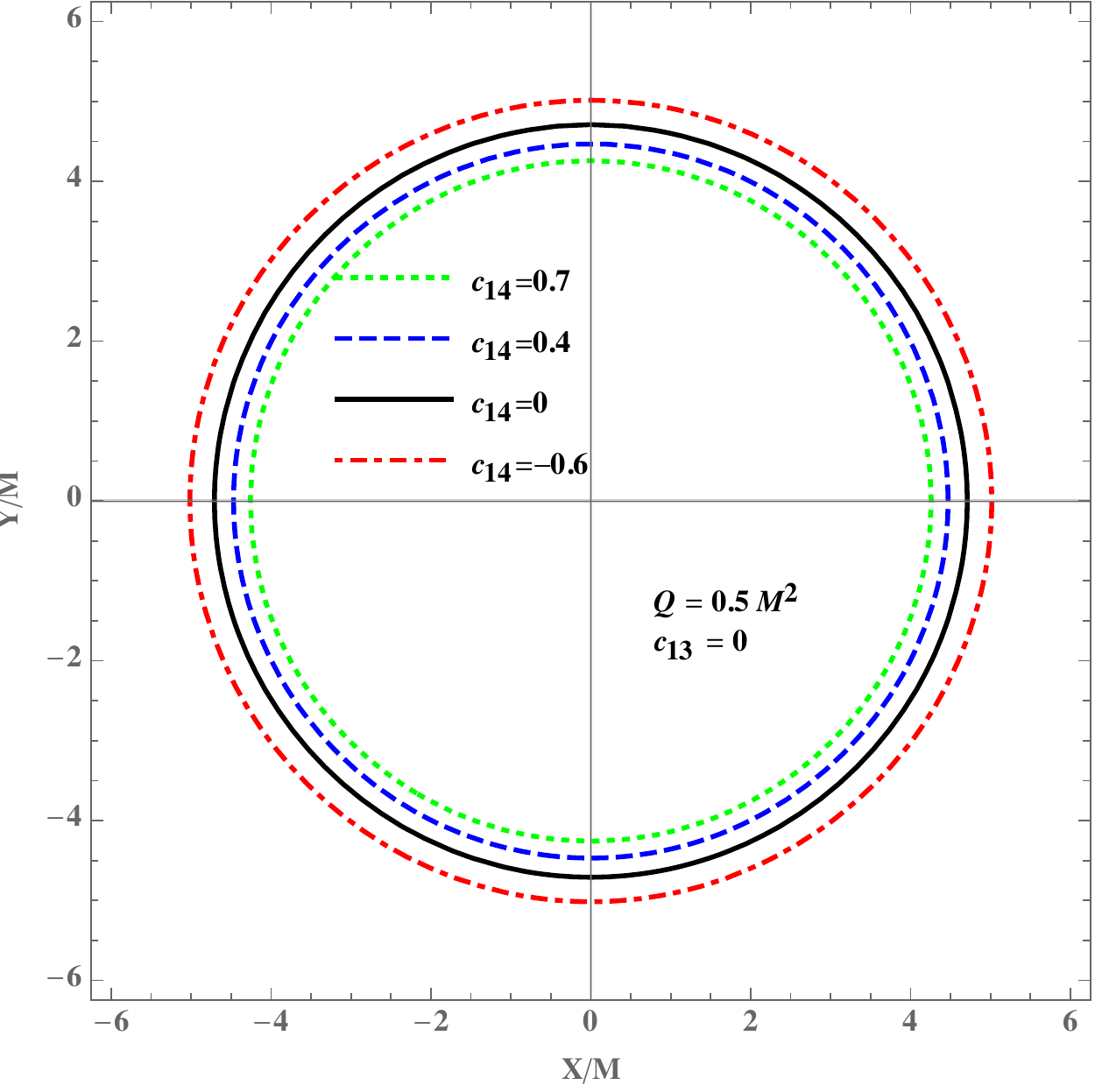}
\includegraphics[width=8.1cm]{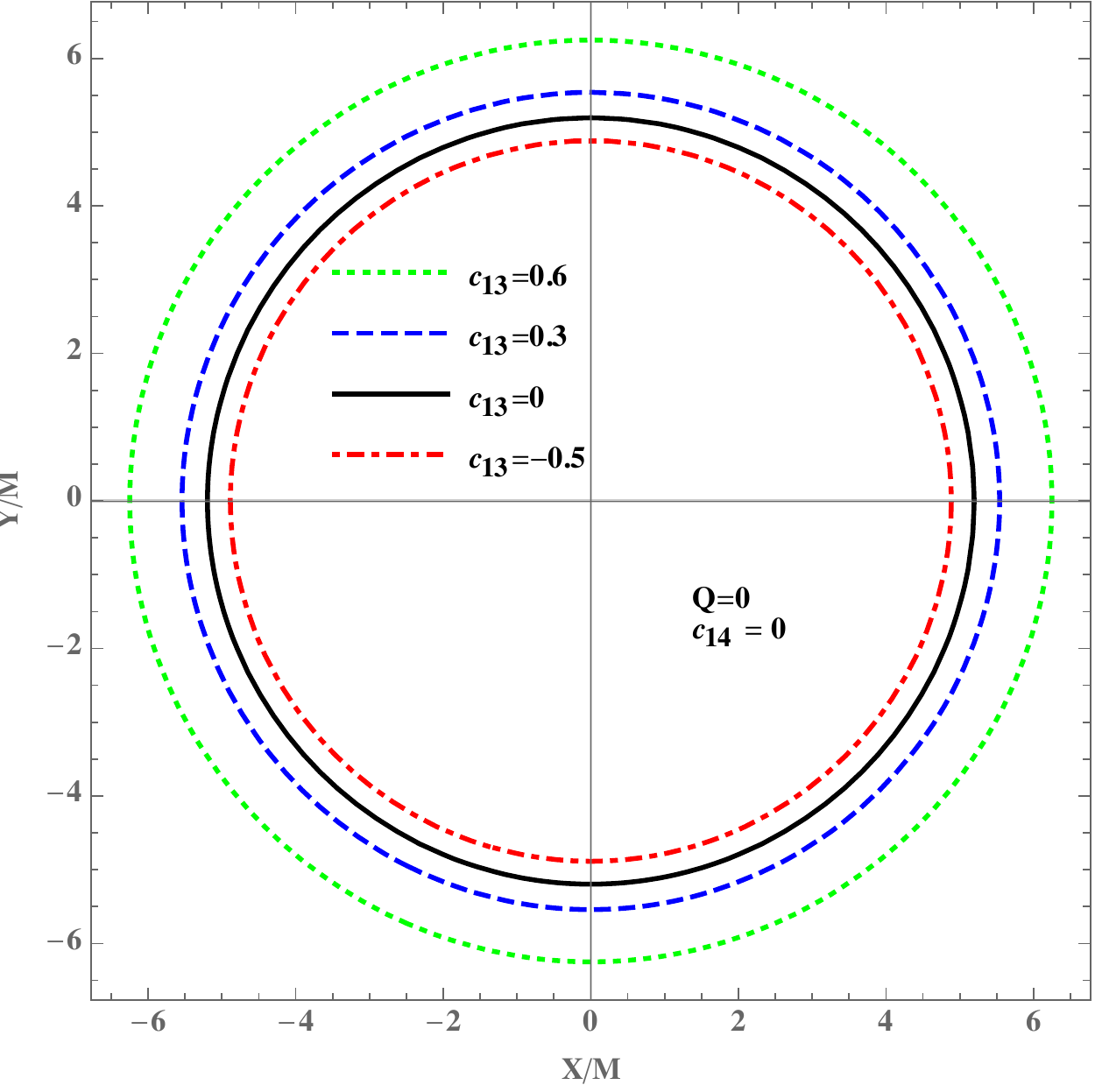}
\includegraphics[width=8.1cm]{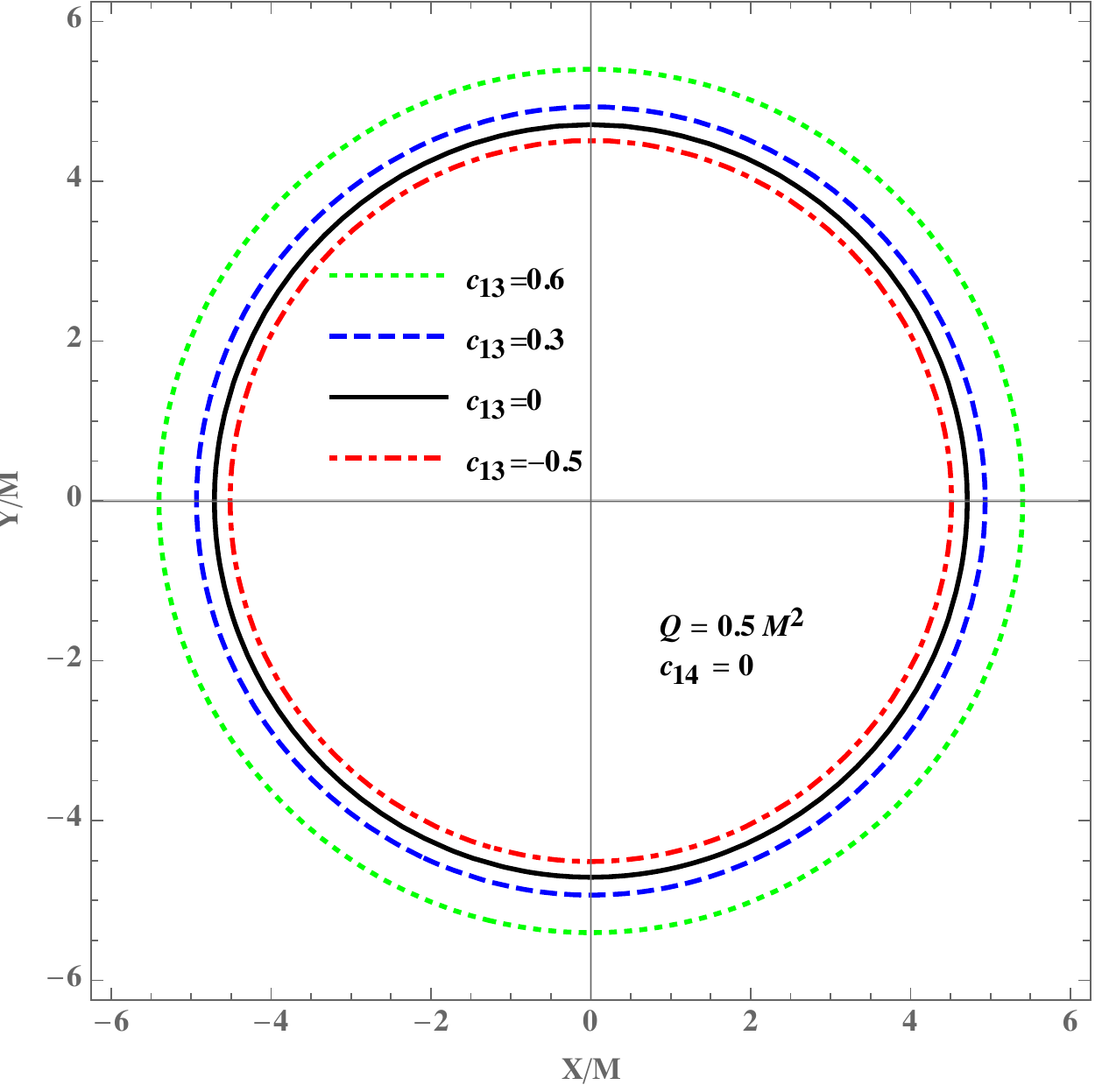}
\caption{Shadow region of the second spherical symmetric Einstein-\ae{}ther black hole ($c_{123}=0$). Left top panel: the neutral case with $Q=0$ and $c_{13}=0$. Right top panel: changed case with $Q=0.5M^2$ and $c_{13}=0$. Left bottom panel: the neutral case with $Q=0$ and $c_{14}=0$. Right bottom panel: changed case with $Q=0.5M^2$ and $c_{14}=0$. } \label{B}
\end{figure*}

The radius of the photon sphere $r_{\rm ps}$ for the two types of the charged and slow-rotating Einstein-\AE{}ther black holes can be determined by the condition (\ref{PSradius}). In the following, we are going to consider the two types of Einstein-\AE{}ther black holes respectively.

\subsection{Black hole shadow with $c_{14}=0$ but $c_{123} \neq 0$}

For the first type charged Einstein-\AE{}ther black hole ($c_{14}=0$ and $c_{123}\neq 0$), this condition leads to a quartic equation,
\bqn
r_{\rm ps}^4 - 3 M r_{\rm ps}^3 + 2 Q r_{\rm ps}^2 - \frac{81 c_{13}}{16(1-c_{13})} M^4=0.
\eqn
Solving this equation, one obtains the radius of the photon sphere,
\bqn\lb{rps1}
r_{\rm ps}&=&\frac{1}{2}\sqrt{\frac{27}{4}M^2-4Q+\frac{27M^3-24M Q}{4 Z_ 2}-Z^2_2} \nb\\
&&+\frac{3M}{4}+\frac{Z_ 2}{2},
\eqn
where
\bqn
Z_ 1& \equiv &\Big[\sqrt{(16 Q^3-144 Q\epsilon+243M^2\epsilon )^2-4(4 Q^2+12\epsilon )^3} \nb\\
    &&    -144 Q\epsilon+243 M^2\epsilon+16 Q^3\Big]^{1/3}, \\
Z_ 2&  \equiv &\sqrt{\frac{4\sqrt[3]{2}\left(3 \epsilon +Q^2\right)}{3 Z_ 1}-\frac{4 Q}{3}+\frac{Z_ 1}{3\sqrt[3]{2}}+\frac{9}{4}M^2},
\eqn
with
\bqn
\epsilon=-\frac{(81 c_{13}) }{16 (1-c_{13})}M^4.
\eqn
It is observe that the radius of the photon sphere can reduce to $r_{\rm ps} = 3 M$ for Schwarzshcild black hole if $c_{13}=0=Q$ and $r_{\rm ps}= 3M(1+\sqrt{1-8Q/(9M^2)})/2$ for RN black hole if $c_{13}=0$. The presence of the \AE{}ther field tends to increase the radius $r_{\rm ps}$ when $c_{13}>0$ and to decrease it if $c_{13} <0$. In Fig.~\ref{A}, we plotted the behavior of shadow by varying the parameter $c_{13}$ for the first type spherical symmetric black hole ($c_{14}=0$ but $c_{123} \neq 0$). We consider the neutral (left panel in Fig.~\ref{A}) and charged (right panel in Fig.~\ref{A}) cases respectively. These figures shows that the shadow region increases with the \ae{}ther parameter $c_{13}$ but shrinks with the electric charge.

The observable $R_{\rm s}$, the size of the shadow of the black holes, can be calculated via
\bqn\lb{RS}
R_{\rm s} = \sqrt{X^2+Y^2} =\frac{r_{\rm ps}}{ \sqrt{e(r_{\rm ps})}}.
\eqn
For the first type spherical symmetric black hole ($c_{14}=0$ but $c_{123} \neq 0$), $r_{\rm ps}$ and $e(r_{\rm ps})$ are given by (\ref{rps1}) and (\ref{e14}) respectively. 

\subsection{Black hole shadow with $c_{123}= 0$}

For the second type of the charged Einstein-\AE{}ther black hole ($c_{123}=0$), from the condition (\ref{PSradius}) we obtain
\bqn\lb{rps2}
r_{\rm ps} = \frac{3M}{2} \left[1+ \sqrt{1- \frac{8(Q/M^2+c_{13}- c_{14}/2)}{9(1-c_{13})}}\right].\nb\\
\eqn
For postive \ae{}ther parameters $c_{13}$ and $c_{14}$, the above expression shows the radius of the shadow region increases with parameter $c_{13}$ but decreases with $c_{14}$. In Fig.~\ref{B}, we display the behaviors of the shadow region for this type black hole by varying the parameter $c_{13}$ and $c_{14}$ respectively. Similar to Fig.~\ref{A}, the neutral (left panel in Fig.~\ref{B}) and charged (right panel in Fig.~\ref{B}) cases are plotted respectively.

The shadow size  $R_{\rm s}$ and its corresponding angular diameters seen from an observer on Earth can be calculated via Eqs. (\ref{RS}) and (\ref{thetas})  with $r_{\rm ps}$ and $(e(r_{\rm ps}))$ given by (\ref{e123}) and (\ref{rps2}) respectively.

\section{Observational constraints}

\begin{figure}
\includegraphics[width=8.1cm]{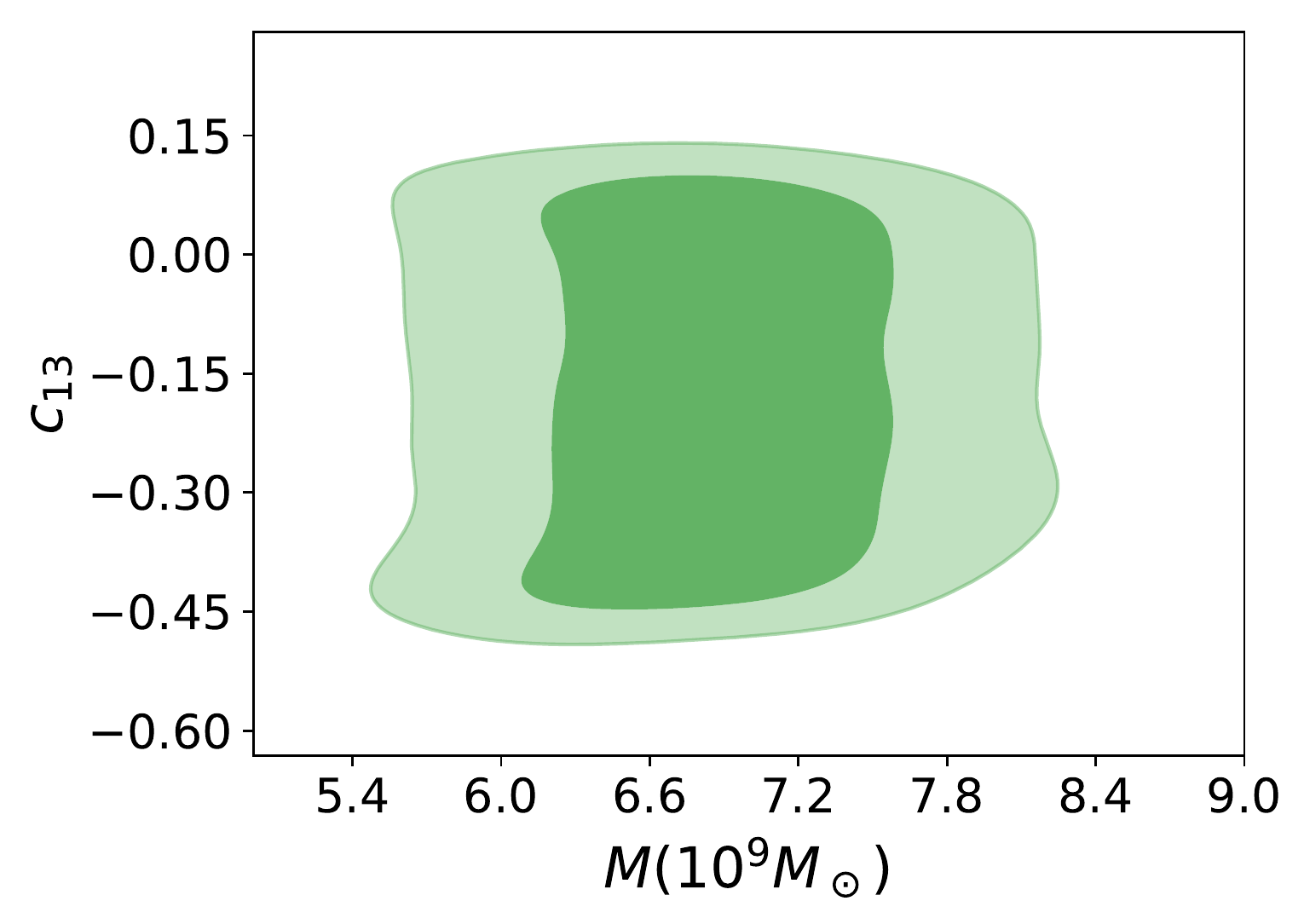}
\includegraphics[width=8.1cm]{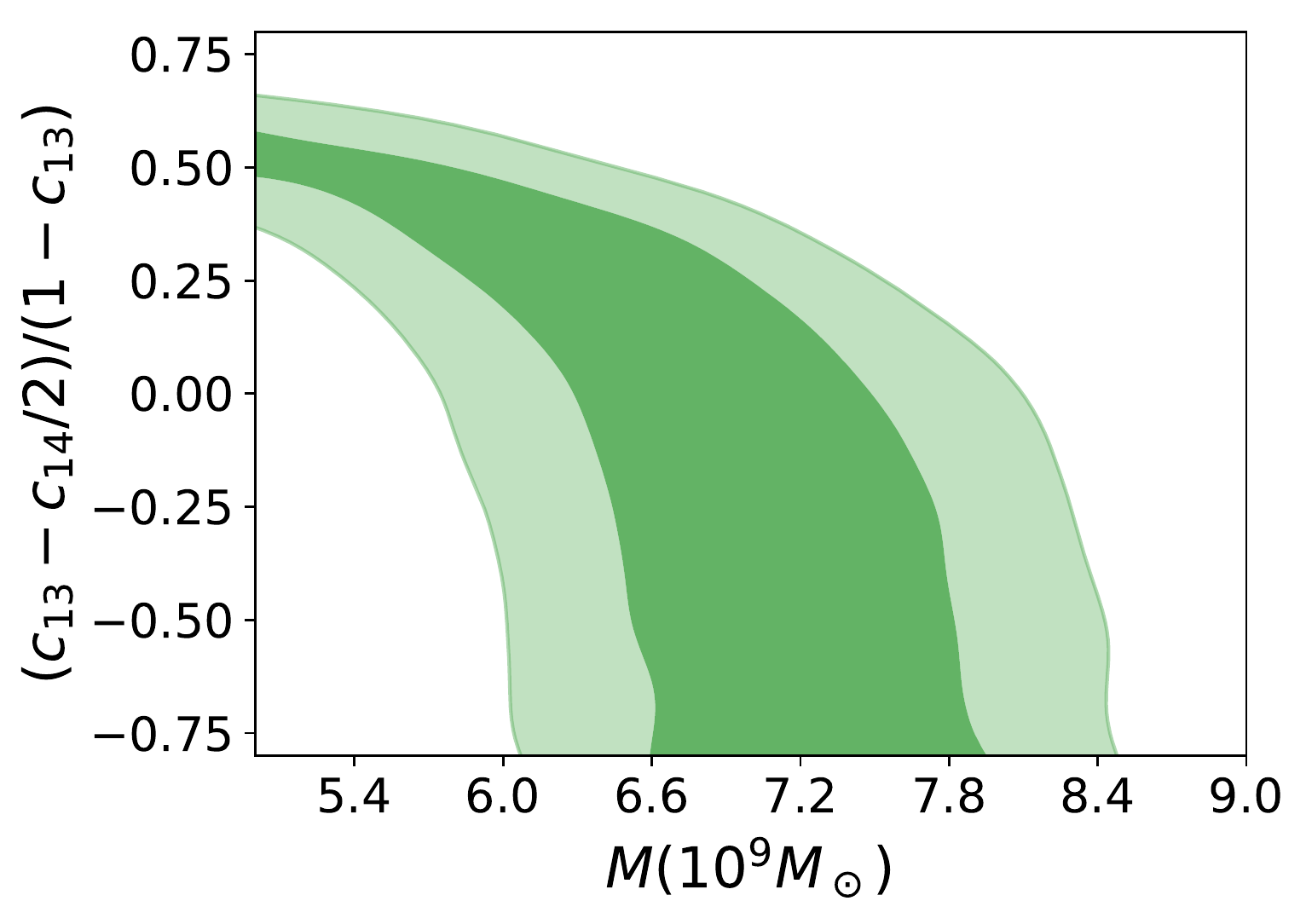}
\caption{68\% and 95\% C.L. constraints on the \ae{}ther parameter and estimated black hole mass $M$ for two types of the Einstein-\AE{}ther black hole. Upper panel: the neutral case with $Q=0$ for the first type black hole. Bottom panel: the neutral case with $Q=0$ for the second type black hole. } \lb{constraint}
\end{figure}

The \ae{}ther parameters $c_{13}$ for the first type black hole and $(c_{13}-c_{14}/2)/(1-c_{13})$ are expected to be constrained by the measurement of the angular diameters of the shadow. For the central supermassive black hole in M87, depending on the distance $D$ between the black hole and the Earth, the angular diameters of the shadow seen by the observer is given by, 
\bqn\lb{thetas}
\theta=2R_s/D,
\eqn
which is $\theta= (42 \pm 3) {\rm \mu{}as}$ as measured from the first image of the black hole by the EHT \cite{m87}. In order to constrain \ae{}ther parameters by using this measurement, we perform the Monte-Carlo simulations for the parameters space $(M,c_{13})$ for the first type black hole and $(M, (c_{13}-c_{14}/2)/(1-c_{13}))$ for the second type black hole. In the simulations, we use the distant $D=(16.8\pm0.8){\rm Mpc}$ \cite{Akiyama:2019eap} and do the simulations for the two types of the black hole respectively. 

In Table.~\ref{table}, we list the 68\% confidence limits for the mass and the \ae{}ther parameters in the Schwarschild Black Hole, the first type Einstein-\AE{}ther black hole, and the second type Einstein-\AE{}ther black hole from observational data of the first image by the EHT, respectively. The constraints on the parameter space $(M,c_{13})$ for the first type black hole and $(M, (c_{13}-c_{14}/2)/(1-c_{13}))$ for the second type black hole are also shown in Fig.~\ref{constraint}. According to these results, the mass of M87 is estimated as $M=(6.8 ^{+0.8}_{-0.7} )\times 10^9M_{\odot}$ in first type black hole and $M=(6.8 \pm 0.6)\times 10^9M_{\odot}$ in second type black hole which are consistent with the value derived by EHT $M=(6.5 \pm 0.7)\times 10^9 M_{\odot}$. The parameter $c_{13}$ in the first type black hole is constrained as
\bqn
c_{13}=-0.18^{+0.27}_{-0.25} \;\;\;  (68\% \;\; {\rm C.L.}), 
\eqn
and the parameter $\frac{c_{13}-c_{14}/2}{1-c_{13}}$ in the second black hole is constrained as 
\bqn
\frac{c_{13}-c_{14}/2}{1-c_{13}} <0.08\;\;\; (68\% \;\; {\rm C.L.}),
\eqn
It is interesting to note that with the presence of the \ae{}ther field, the mass of the central supermassive black hole slightly increases with respect to the case of the Schwarzschild Black Hole for both types of the Einstein-\AE{}ther black holes.

 \begin{table*}
\caption{The 68\% confidence limits for the mass and the \ae{}ther parameters in the Schwarzschild Black Hole, the first type Einstein-\AE{}ther black hole, and the second type Einstein-\AE{}ther black hole from observational data of the first image by the EHT.} \lb{table}
\lb{bestfit}
\begin{ruledtabular}
\begin{tabular}{cccc}
 Parameter  & The Schwarzschild Black Hole & The first type Black Hole & The second type Black Hole  \\
\hline
$M$ ($10^9M_{\odot}$) &  $ 6.5 \pm 0.7$ \cite{m87} &  $ 6.8 \pm 0.6$ & $  6.8^{+0.8}_{-0.7}$ \\
$c_{13}$  & $ ---  $ & $-0.18^{+0.27}_{-0.25}$ & $ ---  $ \\
$\frac{(c_{13}-c_{14}/2)}{(1-c_{13})}$ & $ ---  $  & $ --- $ & $ < 0.08 $  \\
\end{tabular}
\end{ruledtabular}
\end{table*}

%

It is also interesting to compare the above constraints with those from other observations. Recently, the combination of the gravitational wave event GW170817 \cite{abbott_gw170817_2017}, observed by the LIGO/Virgo collaboration, and the event of the gamma-ray burst GRB170817A \cite{abbott_gravitational_2017} provides  a remarkably stringent constraint on the speed of the spin-2 mode, $- 3\times 10^{-15} < c_T -1 < 7\times 10^{-16}$. In the Einstein-aether theory, the speed of the spin-2 graviton is given by $c_{T}^2 = 1/(1-c_{13})$ \cite{jacobson_einstein-eather_2004} with $c_{13} \equiv c_1+c_3$, so the  GW170817 and GRB 170817A events imply
\bq
\lb{2.8a}
\left |c_{13}\right| < 10^{-15}.
 \eq
Together with other observational and theoretical constraints, recently it was found that the parameter space of the theory is further restricted to \cite{aw}
 \bq
\lb{2.8b}
c_{4} \lesssim 0, \quad 0 \lesssim c_2 \lesssim 0.095, \quad 0 \lesssim c_{14} \lesssim 2.5\times 10^{-5}.
 \eq
Obviously, these constrains are much stringent than those obtained from the first image of the EHT. Although the observational constraints from the first black hole image is not accurate enough with respect to other type observations, our results show that the observational data from the images by EHT does have the capacity for constraining black hole parameters beyond those in GR. In particular, the two types of the Einstein-\AE{}ther black holes are obtained for several special combination of the \ae{}ther parameters. In general, the Einstein-\AE{}ther black hole can contain more \ae{}ther parameters, such as $c_3$ or $c_{123}$ which are less constrained by the observations. We expect that the measurement of the first black hole image and future precise observations can provide more significant information on the constraints of these parameters, which will be studied in our future works.

\section{Deflection angle}
\renewcommand{\theequation}{5.\arabic{equation}} \setcounter{equation}{0}

Here we shall compute the deflection angle of light in the spacetime metric (\ref{slow1}) with $e(r)$ given by (\ref{e123}) and (\ref{e14}) for the two types of the Einstein-\AE{}ther black holes respectively. To do so, we shall use a recent geometric method (also known as Gibbons-Werner's method) based on the application of the Gauss-Bonnet theorem (GBT) over the optical geometry \cite{gibbons}. Note that the deflection angle for static spacetimes were computed in Ref.~\cite{gibbons}. Werner extended this method to compute the deflection angle by a Kerr black hole using the Kerr-Randers optical geometry \cite{werner}. Furthermore Ishihara et al. \cite{ishihara1,ishihara2} introduced another approach to compute the finite distance corrections in Kerr spacetime. 

Here, we shall show that for a slowly rotating metric at the linear order of its spin $a$, the deflection angle can be evaluated in a rather simple way.  Toward this purpose, let us rewrite the metric (\ref{slow1}) in the equatorial plane ($\theta=\pi/2$) in the following form,
\begin{equation}
ds^2 = -\left(e(r)+\frac{4 M a}{r}\frac{d\phi}{dt} \right)dt^2+\frac{dr^2}{e(r)}+r^2d\phi^2.
\end{equation}
Since the quantity $d\phi/dt$ can be evaluated from the Lagrangian of the particle later, it is evident that this reduces the problem similar to that in \cite{gibbons}. 
Note that other similar cases were also considered in Refs. \cite{Jusufi:2018kry,Jusufi:2018gnz} to compute the deflection of massive particles/light in a linearized Kerr and Kerr-like metrics. To proceed for the later calculation, it is convenient to introduce two new variables $dr_\star$ and $f(f(r_\star))$ in the form
\begin{eqnarray}
dr_{\star}&=&\frac{dr}{\sqrt{e(r) \left(e(r)+\frac{4 M a}{r}\frac{d\phi}{dt}\right)}},\\
f(r_{\star})&=& \frac{r}{\sqrt{e(r)+\frac{4 M a}{r}\frac{d\phi}{dt}}}.
\end{eqnarray}
Then the optical metric reduces to the following form
\begin{equation}
dt^2 ={dr_{\star}}^2+f^2(r_{\star})d\phi^2.
\end{equation}
In what follows we shall consider the two types of the Einstein-\AE{}ther black holes individually to evaluate the deflection angle.

\subsubsection{Case $c_{123}=0$}

Let us first consider the case with $c_{123}=0$ and start with the Lagrangian of the particle in a curved spacetime, which is given by
\begin{equation}
\mathcal{L}=\frac{1}{2}g_{\mu \nu}\dot{x}^{\mu}\dot{x}^{\nu}.
\end{equation}
From the spacetime symmetries we can obtain two constants of motion, namely the conservation of energy, $E$, and angular momentum $J$. Therefore, from (\ref{slow1}) it follows that
\begin{equation}\label{ConservationEL}
\begin{split}
    p_{t}&=-e(r)\dot{t}-\frac{2Ma}{r}\dot{\phi}=-E, \\
    p_{\phi}&=r^2 \dot{\phi}-\frac{2Ma}{r}\dot{t}=L.
\end{split}
\end{equation}
where dot represents derivation with respect to affine parameter $\lambda$, and $e(r)$ is given by Eq. (\ref{e123}). Combining these equations we can find the quantity $d\phi/dt$, which is given by
\begin{equation}\label{TimeAzimuthalEq}
\frac{d\phi}{dt}=\frac{b (r-2M)+\frac{Q^2b}{(1-c_{13})r}-\frac{2 c_{13}-c_{14}}{2(1-c_{13})}\frac{M^2 b}{r} +2Ma}{r^3-2Ma b}.
\end{equation}
Note that we have used the energy $E$ and the angular momentum of the particle (photon) $J$ to define the impact parameter $b$ at infinity via
\begin{equation}
b=\frac{E}{L}.
\end{equation}

With these results in hand, we can apply the Gauss-Bonnet theorem (GBT). Let $\mathcal{D}_{R}$ be a non-singular optical region with boundary $\partial
\mathcal{D}_{R}=\gamma _{\tilde{g}}\cup C_{R}$. The GBT simply relates the optical geometry (for example, the optical curvature) with the topology (usually characterized by the Euler characteristic number) in terms of the following relation
\begin{equation}
\iint\limits_{\mathcal{D}_{R}}\mathcal{K}_0\,\mathrm{d} \mathcal{S}+\oint\limits_{\partial \mathcal{%
D}_{R}}\kappa \,\mathrm{d}t+\sum_{i}\theta _{i}=2\pi \chi (\mathcal{D}_{R}),
\end{equation}
with $\kappa$ being the geodesic curvature, while $\mathcal{K}_0$ is known as the Gaussian optical curvature. 
Now we need to calculate this two quantities. Firstly, the Gaussian optical curvature $\mathcal{K}_0$ can be computed from the following definition \cite{gibbons},
\begin{equation}
  \mathcal{K}_0 =   - \frac{1}{f (r_{\star})}  \left[ \frac{\mathrm{d} r}{\mathrm{d}
  r_{\star}}  \frac{\mathrm{d}}{\mathrm{d} r} \left( \frac{\mathrm{d}
  r}{\mathrm{d} r_{\star}} \right) \frac{\mathrm{d} f}{\mathrm{d}r} + \left(
  \frac{\mathrm{d}r}{\mathrm{d} r_{\star}} \right)^2 \frac{\mathrm{d}^2
  f}{\mathrm{d} r^2} \right]. 
\end{equation}
Applying in our case, we can find the Gaussian optical curvature at the leading order of $a$ as follows,
\begin{align}
\begin{split}
\mathcal{K}_0 \simeq  - \frac{2M}{r^3}+\frac{3 Q^2 \left(1+c_{13}\right)}{r^4}+\frac{18 M ab}{r^5}.\lb{4.28}
\end{split}
\end{align}

Secondly, for the geodesic curvature $\kappa$ of the optical geometry, it is defined by
\begin{equation}
\kappa =\tilde{g}\,\left(\nabla _{\dot{\gamma}}\dot{\gamma},\ddot{\gamma}\right)
\end{equation}
with condition (unit speed condition)  $\tilde{g}(\dot{\gamma},\dot{%
\gamma})=1$, where $\ddot{\gamma}$ represents the unit acceleration vector. It is evident that there is a zero contribution from the geodesics i.e. $\kappa (\gamma_{\tilde{g}})=0$, hence the only contribution is due to the curve $C_{R}$
\begin{equation}
\kappa (C_{R})=|\nabla _{\dot{C}_{R}}\dot{C}_{R}|.
\end{equation}
To compute this quantity we need to keep in mind that for large and constant radial coordinate 
$C_{R}:=r(\phi)=R=\text{const}$, the nonzero (radial) contribution implies
\begin{equation}
\left( \nabla _{\dot{C}_{R}}\dot{C}_{R}\right) ^{r}=\dot{C}_{R}^{\phi
}\,\left( \partial _{\phi }\dot{C}_{R}^{r}\right) +\tilde{\Gamma} _{\phi
\phi }^{r}\left( \dot{C}_{R}^{\phi }\right) ^{2}. 
\end{equation}
While the first terms vanishes, using the unit speed condition we find that
\begin{eqnarray}\lb{4.32}
\lim_{R\rightarrow \infty }\kappa (C_{R}) &=&\lim_{R\rightarrow \infty
}\left\vert \nabla _{\dot{C}_{R}}\dot{C}_{R}\right\vert \rightarrow \frac{1}{R}. 
\end{eqnarray}
Note that for an observer located at a very large distance we also have 
\begin{eqnarray}\lb{4.33}
\lim_{R\rightarrow \infty } \frac{dt}{d\phi} &\to &  R .
\end{eqnarray}%

From the GBT we need to express the deflection angle, by choosing a non-singular domain which has  Euler characteristic number 
$\chi (\mathcal{D}_{R})=1$. By construction, there are two corresponding jump angles in the limit $R\rightarrow \infty$, namely the following condition is recovered $\theta _{\mathit{O}%
}+\theta _{\mathit{S}}\rightarrow \pi $ \cite{gibbons}. The GBT can be rewritten as follows
\begin{equation}
\lim_{R\rightarrow \infty }\int_{0}^{\pi+\hat{\alpha}}\left[\kappa \frac{\mathrm{d} t}{\mathrm{d} \phi }\right]_{C_R} \mathrm{d} \phi =\pi-\lim_{R\rightarrow \infty }\iint\limits_{\mathcal{D}_{R}}\mathcal{K}\,\mathrm{d}\mathcal{S}.
\end{equation}

Using Eqs. (\ref{4.32}, \ref{4.33}) and the Gaussion optical curvature (\ref{4.28}), the deflection angle  can be computed via
\begin{equation}
\hat{\alpha}=-\int\limits_{0}^{\pi}\int\limits_{\frac{b}{\sin \phi}}^{\infty}\left(- \frac{2M}{r^3}+\frac{3 Q^2 \left(1+c_{13}\right)}{r^4}+\frac{18 M ab}{r^5}\right)d\mathcal{S},
\end{equation}
where $dS=\sqrt{\tilde{g}} \,dr_{\star} d\phi\simeq r dr d\phi$, is the surface element of the optical geometry. It is worth noting that in the weak limit, we have used the light ray equation $r=b/\sin \phi$ in the integration domain. Evaluating the last integral is not difficult to show that 
\begin{equation}
\hat{\alpha}\simeq \frac{4M}{b}-\frac{3\pi Q^2\left(1+c_{13}\right)}{4b^2}\pm \frac{4Ma}{b^2}.
\end{equation}

This result shows that the deflection angle at the leading order of $a$ depends on the parameter $c_{13}$. Letting $c_{13}=0$, we do recover the deflection angle by a the Kerr-Newmann black hole. Let us also mention that the positive and negative sign is for a prograde and retrograde light ray, respectively. Namely, $a>0$ means that BH is co-rotating relative to the observer, while for $a<0$ it is counter-rotating relative to BH. 

In Fig.~\ref{fig6} we plot the deflection angle against the impact parameter and black hole spin considering both prograde (or direct) and retrograde photon orbits. These orbits are generally defined by the photon motion in the direct or opposite allignment with the spin of black hole. The left panel shows that the bending angle increases for photons approaching the black hole in retrograde orbits with smaller impact parameter. This behavior is similar to the case for Kerr black hole lensing as well \cite{iyer}. However for prograde orbits, the bending angle first approaches to a maximum for $b\approx 2$ and than sharply gets negative. The negative bending angle suggests that the photons are orbiting the black hole with negative angle (clockwise motion).  The right panel shows a linear increase (decrease) of the bending angle relative to the black hole spin for prograde and retrograde orbits.

\subsubsection{Case $c_{14}=0$ and $c_{123} \neq 0$}
Following the same method and using the spacetime metric (\ref{slow1}) with $e(r)$ given by (\ref{e14}) for the quantity $d\phi/dt$, we find
\begin{equation}
\frac{d\phi}{dt}=\frac{b (r-2M)+\frac{Q^2b}{r}-\frac{27 c_{13}}{256(1-c_{13})}\frac{(2M)^4 b}{r^3} +2Ma}{r^3-2Ma b}.
\end{equation}
Then the Gaussian optical curvature $\mathcal{K}_0$ gives
\begin{align}
\begin{split}
\mathcal{K}_0 \simeq - \frac{2M}{r^3}+\frac{3 Q^2}{r^4}+\frac{18 M ab}{r^5}.
\end{split}
\end{align}
It is obvious at the leading order of $a$, the \ae{}ther parameters have no effects on $\mathcal{K}_0$. Then the deflection angle for this type of black hole can be computed via
\begin{equation}
\hat{\alpha}=-\int\limits_{0}^{\pi}\int\limits_{\frac{b}{\sin \phi}}^{\infty}\left(- \frac{2M}{r^3}+\frac{3 Q^2 }{r^4}+\frac{18 M ab}{r^5}\right)d\mathcal{S},
\end{equation}
which yields
\begin{equation}
\hat{\alpha}\simeq \frac{4M}{b}-\frac{3\pi Q^2}{4b^2}\pm \frac{4Ma}{b^2}.
\end{equation}
This result is nothing else but the deflection angle in the Kerr-Newmann geometry at the linear order of $a$. Thus, the deflection angle at the linear order is not affected by the \ae{}ther parameter $c_{13}$. Of course, one can consider higher order terms combined with the appropriate equation of light ray in the integration domain to find higher order correction terms, but this is beyond the scope of the present paper.

\begin{figure*}
\includegraphics[width=8.1cm]{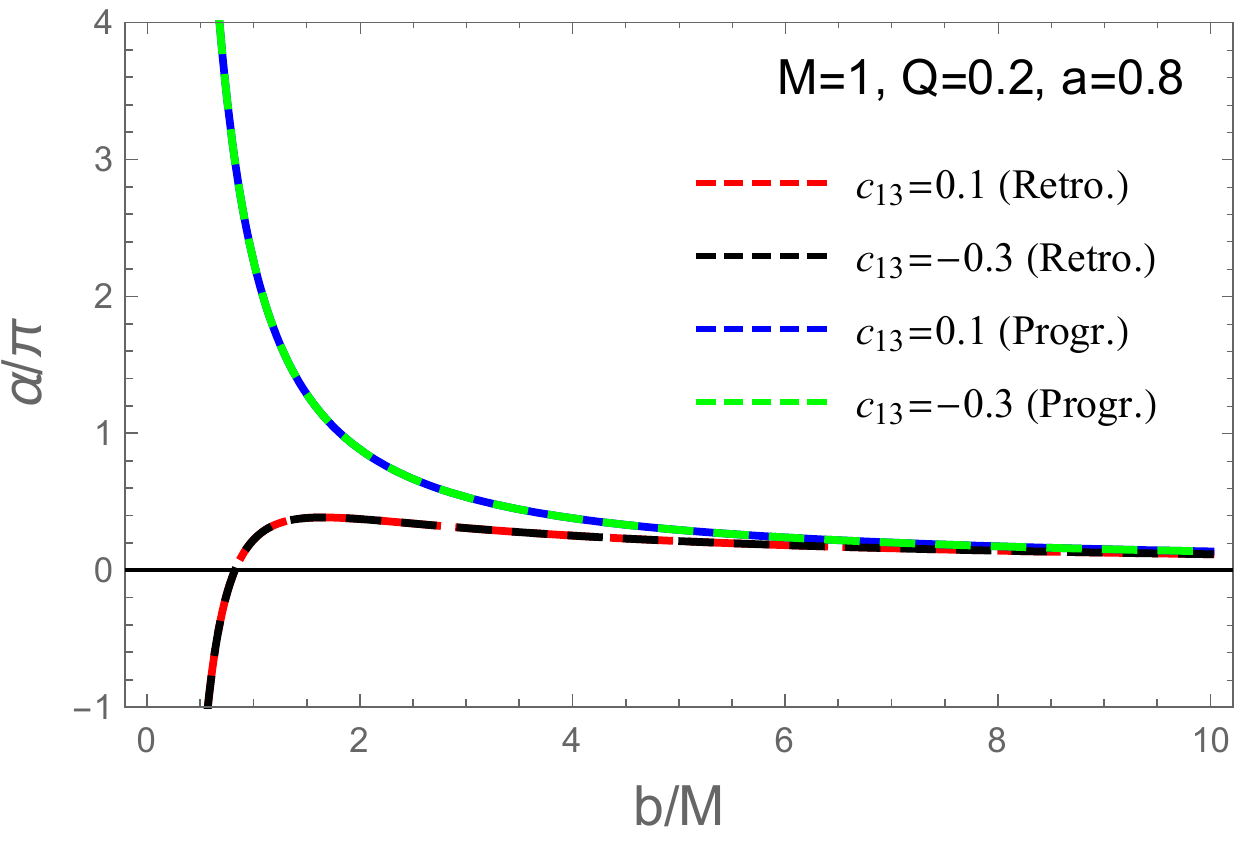}
\includegraphics[width=8.1cm]{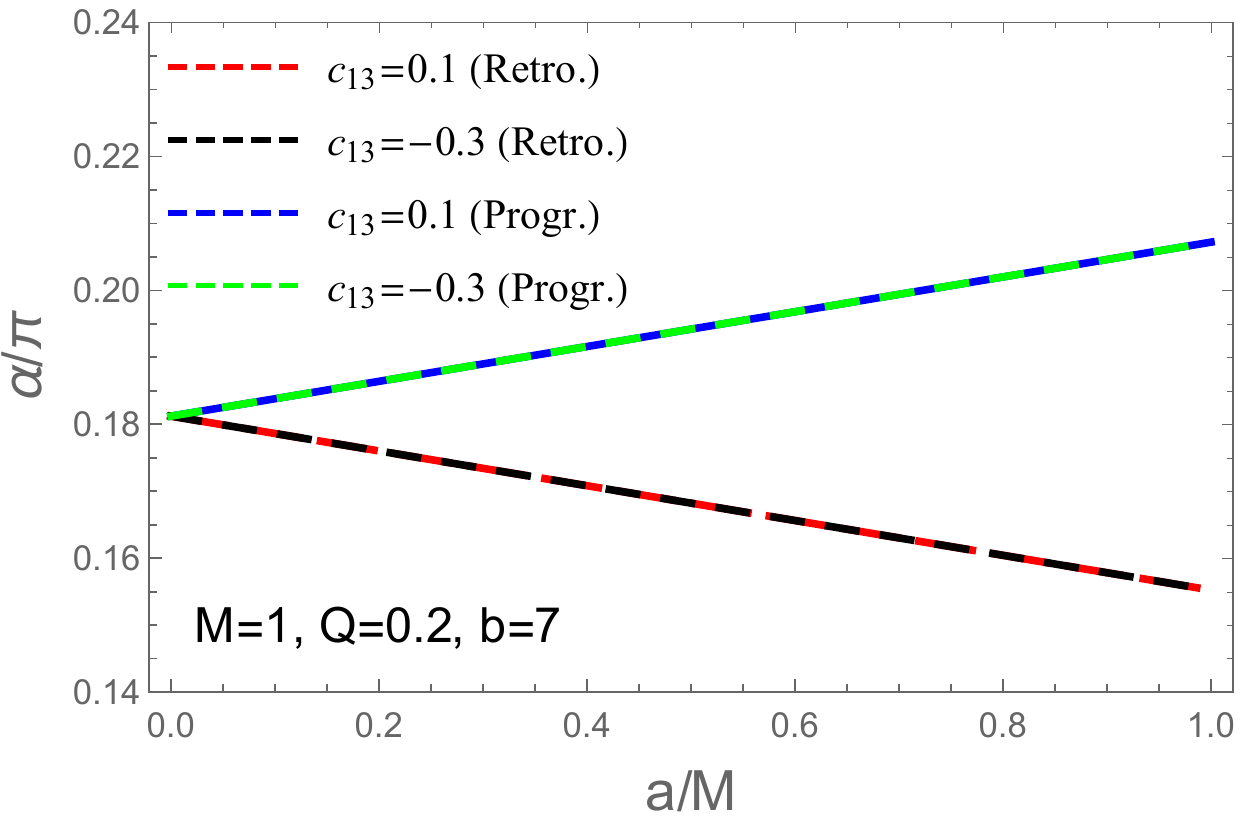}
\caption{Left panel: The deflection angle as a function of the impact parameter for the second type of the Einstein-\AE{}ther black hole. We observe that the effect of the \ae{}ther parameter on the deflection angle are very small.  Right panel: Deflection angle as a function of the angular momentum parameter for constant $b$. Here it is shown that deflection angle linearly increases (decreases) with the increase of $a$. Overall, as expected, the effect of \ae{}ther parameter in the deflection angle are incredibly small, which are beyond the scope of the present day technology.}\lb{fig6}
\end{figure*}

\section{Time Delay}
\renewcommand{\theequation}{6.\arabic{equation}} \setcounter{equation}{0}

In this section, we analyze the time delay in a gravitational field of the black hole solutions given by (\ref{slow1}) for two types of the Einstein-\AE{}ther black holes. Normally, if two photons are emitted at the same time but follow different paths to reach the observer, then they will take two different times to reach the observer and this time difference is called the \textit{time delay}. To proceed, let us rewrite (\ref{slow1}) line element in the following form
\begin{equation}
ds^2 = -A(r) dt^2 + B(r) dr^2 + C(r)(d\theta^2+\sin^2\theta d\phi^2),
\end{equation}
where 
\begin{eqnarray}
A(r)&=&e(r)+\frac{4Ma}{r}\sin^2\theta\frac{d\phi}{dt},\\
B(r)&=&e(r)^{-1},\,\,\, C(r)=r^2.
\end{eqnarray}
For simplicity let us consider $\theta=\pi/2$, then the total time required for a light signal passing through the gravitational field of the black hole to go from the observer (Earth) to the source and back after reflection from the source is given by \cite{weinberg1}, 
\bqn\lb{timedelay}
\Delta T &=& 2\int_{r_{0}}^{r_{e}} \mathcal{I}(r,r_0)dr + 2\int_{r_{0}}^{r_{s}} \mathcal{I}(r,r_0)dr,
\eqn
where $r _{e} $ and $r _{s} $ are distances of the observer (Earth) and the source from the massive body respectively, $r _{0} $ is the closest approach to the black hole, and
\bqn
\mathcal{I}(r,r_0) &=& \left(\frac{A(r)}{B(r)}-\frac{A^2(r)C(r_0)}{B(r)C(r)A(r_0)}\right)^{-\frac{1}{2}}-\left(1-\frac{r_0^2}{r^2}\right)^{-\frac{1}{2}}.\nb
\eqn
With above relation, we can calculate the time delay due to the massive gravitational field of the black hole. In the following we will consider both types of Einstein-\AE{}ther black holes individually.

\subsubsection{Case $c_{123}=0$}
Let us consider the case with $e(r)$ given by Eq. (\ref{e123}) in this subsection. Considering $r _{e} $,$r _{s} $, $r _{0} $ $ \gg  2M$ and evaluating the integrals in (\ref{timedelay}), we can estimate the time delay due to the gravitational field of the  black hole as
\begin{eqnarray}\lb{time}
\Delta T _{e},_{s}&=& 4M\ln\left[\frac{\left(r_{e}+\sqrt{r_{e}^{2}-r_{0}^{2}}\right)\left(r_{s}+\sqrt{r_{s}^{2}-r_{0}^{2}}\right)}{r_{0}^{2}}\right]\nb\\
&+&2M\left(\sqrt{\frac{r_{e}-r_{0}}{r_{e}+r_{0}}}+\sqrt{\frac{r_{s}-r_{0}}{r_{s}+r_{0}}}\right)\nb\\
&+& \frac{3 Q^2(1+c_{13} )}{r_0}\left(\arcsin\frac{r_0}{r_s}+\arcsin\frac{r_0}{r_e}-\pi \right)\nb\\
&-& \frac{4M a}{r_0}\Bigg[\left(3+\frac{2r_0}{r_e}\right)\sqrt{\frac{r_{e}-r_{0}}{r_{e}+r_{0}}} \nb\\
&&~~~~~~~~~+\left(3+\frac{2r_0}{r_s}\right)\sqrt{\frac{r_{s}-r_{0}}{r_{s}+r_{0}}}\Bigg].
\end{eqnarray}
This shows that, time delay is affect by $c_{13}$ simply by replacing the charge $Q^2 \to Q^2(1+c_{13})$.

\subsubsection{Case $c_{14}=0$ and $c_{123} \neq 0$}

In this case the \ae{}ther parameter $c_{13}$ does not have effects on the time delay, at least to the degree of approximation adopted here. Due to the economy of space, without going into details the total time delay is given by Eq. (\ref{time}) with $c_{13}=0$.

\section{Conclusions and Outlook}
\renewcommand{\theequation}{5.\arabic{equation}} \setcounter{equation}{0}

In this paper, we study the shadow cast by two types of charged and slowly rotating black holes in the Einstein-\AE{}ther theory of gravity. By using the null geodesics of these two types of the black holes, we derive analytically the apparent shadow size which reflects both the effects of the charge and the \ae{}ther field. It is shown that with the presence of the \ae{}ther field, the shadow size of the black hole increases with the parameter $c_{13}$ in the both types of the black holes but decreases with the parameter $c_{14}$ in the second type of black hole. In addition, with the analytical expressions of the apparent size of the shadow, we perform the Monte-Carlo simulations for the parameters space $(M,c_{13})$ for the first type black hole and $(M, (c_{13}-c_{14}/2)/(1-c_{13}))$ for the second type black hole by using the measurement of the angular diameters of the shadow from the first black hole image of M87. The results of these constraints are presented in Table.~\ref{table} and Fig.~\ref{constraint}. 

Furthermore, we have investigated the effect of \ae{}ther field on the deflection angle of light and time delay. We have argued that there is a tiny effect of the parameter $c_{13}$ on the deflection angle only in the second type of black hole, $c_{123} =0$. As can be seen, this effect is encoded by replacing $Q^2 \to Q^2(1+c_{13})$. However, we find no effect on the deflection angle/time delay for the first type of black hole, $c_{14}=0$ but $c_{123} \neq 0$.  At last, our analyses is valid in leading order terms. 

With the above main results, we would like to mention several directions that can be carried out to extend our analysis.  First, in order to study the photon trajectories, we have assumed that it follows null geodesics. However, this is only correct if there is no coupling between the electromagnetic field and the \ae{}ther field. If the interaction between the photon and the \ae{}ther field exists, this interaction can affect the photon trajectories around the black hole that deviates from the null geodesics. {This leads two effects. First, such interaction can slightly modify both the shape and the size of the black hole shadow, the second, it could also lead to significant modifications during the propagating of the light to a distant observer.} It is interesting to explore whether such interaction can affect the shadow size and how the future observations can constrain the interaction. 

Second, the two types of the black holes in the Einstein-\AE{}ther black hole we studied in this paper are corresponding to two specific combinations of the coupling constants of the \ae{}ther field, i.e., $c_{14}=0$ but $c_{123} \neq 0$ for the first type black hole and $c_{123} =0$ for the second type black hole, respectively. This is because this two types of black holes are the only two analytical black hole solutions so far in the Einstein-\AE{}ther theory. But in general one can construct black hole solution numerically, thus it is also interesting to study the effects of the \ae{}ther field on the black hole shadow and their observational constraints with more \ae{}ther parameters. {Alternatively, another approach is to construct a analytical representations based on the numerical solution, as shown in \cite{Eling:2006ec, Konoplya:2006rv, Konoplya:2006ar}. Recently, some analytical approximate space-time solutions have been constructed by using the continued-faction expansion in the Einstein-Scalar-Gauss-Bonnet gravity and Einstein-Scalar-Maxwell gravity with scalarization \cite{Konoplya:2019fpy, Konoplya:2019goy}. The shadow of the approximate black holes have also been explored. It is also interesting to construct analytical approximate space-time solution by using similar expansion and study their shadow properties. This will be considered in our future works.}

At last, in our study we have performed the Monte-Carlo simulations for the parameters space of parameters in the black hole spacetime by using the measurement of the angular diameters of the shadow from the first black hole image of M87. It would be very interesting to extend this analysis to other types of black holes with more precise observational data in the future. 

We hope to return to the above issues soon in future studies to possibly extend some of these results.


\section*{Acknowledgements}
This work is supported by the National Natural Science Foundation of China with the Grants Nos. 11675143 (T.Z. \& Q.W.) and the Fundamental Research Funds for the Provincial Universities of Zhejiang in China with Grants No. RF-A2019015 (T.Z. \& Q.W.).

\end{document}